\documentclass[letterpaper]{article} 
\usepackage[draft]{aaai25}  
\usepackage{times}  
\usepackage{helvet}  
\usepackage{courier}  
\usepackage[hyphens]{url}  
\usepackage{graphicx} 
\usepackage{natbib}  
\usepackage{caption} 
\usepackage{algorithm}
\usepackage{algorithmic}
\usepackage{amsmath,graphicx}
\usepackage{float}
\usepackage{listings}
\usepackage{mathtools}
\usepackage{upgreek}
\usepackage{tabularx}
\usepackage{siunitx}
\usepackage{caption}
\usepackage{enumitem}
\usepackage{multirow}
\usepackage{pifont}
\usepackage{multicol}
\usepackage{booktabs}
\usepackage{subcaption}
\usepackage{amsfonts}
\usepackage{amssymb}

\usepackage{etoolbox}

\AtBeginEnvironment{tabular}{\scriptsize}

\usepackage{newfloat}
\usepackage{listings}
\DeclareCaptionStyle{ruled}{labelfont=normalfont,labelsep=colon,strut=off} 
\floatstyle{ruled}
\newfloat{listing}{tb}{lst}{}
\floatname{listing}{Listing}
\title{MuMu-LLaMA: Multi-modal Music Understanding\\and Generation via Large Language Models}

\author {
    Shansong Liu\textsuperscript{\rm 1}\footnote{\noindent Equal contribution.}\footnote{\noindent Corresponding author.\\ \hspace*{1.2em} This paper presents an updated version of M$^{2}$UGen by the same authors. Please reference this version in future citations.},
    Atin Sakkeer Hussain\textsuperscript{\rm 2}\footnotemark[1],
    Qilong Wu\textsuperscript{\rm 2}\footnotemark[1],
    Chenshuo Sun\textsuperscript{\rm 2},
    Ying Shan\textsuperscript{\rm 1}
}
\affiliations {
    \textsuperscript{\rm 1}ARC Lab, Tencent PCG\\
    \textsuperscript{\rm 2}National University of Singapore\\
    {\tt\small dadinghh2@gmail.com}, {\tt\small atin.s@u.nus.edu}, {\tt\small qilong\_wu@u.nus.edu}, 
    \\{\tt\small csun@nus.edu.sg}, {\tt\small yingsshan@tencent.com}\\
    \small{Code:~\texttt{https://github.com/shansongliu/MuMu-LLaMA}}
}

\usepackage{bibentry}

\begin{document}

\maketitle

\begin{figure*}[!t]
    \centering
    \includegraphics[width=15cm]{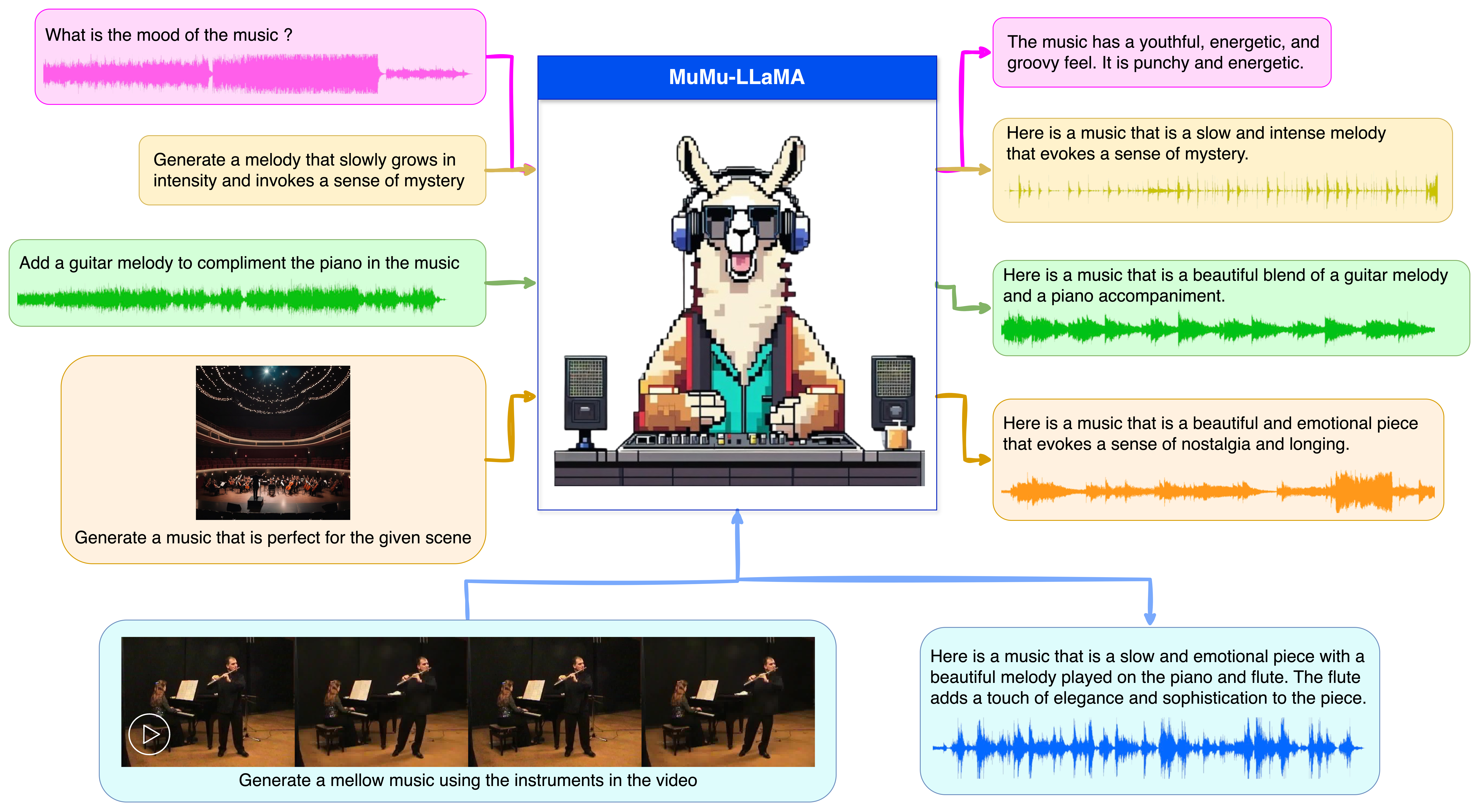}
    \caption{Multi-modal music understanding and generation by our proposed MuMu-LLaMA framework.}
  \label{fig:cover}
\end{figure*}

%

\small
\begin{abstract}


The landscape of research leveraging large language models (LLMs) has seen remarkable growth, with numerous studies harnessing these models’ powerful reasoning capabilities across modalities like text, speech, images, and videos. However, the domain of multi-modal music comprehension and generation remains relatively unexplored, primarily due to the lack of a comprehensive well-annotated multi-modal music dataset. To address this gap, we introduce a novel dataset containing 167.69 hours of multi-modal data, including text, images, videos, and music annotations, tailored for multi-modal music understanding and generation, annotated using advanced visual models like LLaVA and Video-LLaVA. Based on this well-annotated dataset, we propose a multi-modal music understanding and generation model named MuMu-LLaMA. This framework integrates LLMs to comprehend input music and generate music across various modalities, utilizing pretrained models for music, images, and videos. For music generation, we incorporate AudioLDM 2 and MusicGen, connecting multi-modal understanding with music generation through the LLaMA model. Our comprehensive evaluation, encompassing four key tasks—music understanding, text-to-music generation, prompt-based music editing, and multi-modal music generation—demonstrates that MuMu-LLaMA outperforms current state-of-the-art models, highlighting the potential of combining LLMs with multi-modal inputs for innovative music applications.

\end{abstract}

\section{Introduction}
\label{sec:intro}

The landscape of research leveraging large language models (LLMs) has seen remarkable growth, with numerous studies harnessing these models' powerful reasoning capabilities across modalities like text, speech, images, and videos. These models facilitate semantic comprehension and interaction within and across modalities, enabling dynamic conversations \cite{openai_chatgpt,touvron2023llama}, sophisticated audio and video event recognition \cite{tang2023salmonn}, and detailed image and 3D data annotation \cite{xu2023pointllm}. Despite these advancements, the domain of multi-modal music comprehension and generation remains relatively unexplored, primarily due to the lack of a comprehensive, well-annotated multi-modal music-centric dataset, which is crucial for instruction tuning in this domain.

Multi-modal large language models (MLLMs) have emerged as a thriving area of research, captivating the current scholarly landscape \cite{yin2023survey}. They primarily serve as a bridge connecting diverse modalities, such as visual \cite{alayrac2022flamingo,li2023blip,pmlr-v209-xu23a}, audio \cite{tang2023salmonn,huang2023audiogpt,liu2023music}, 3D \cite{xu2023pointllm,sun20233d} and so on, transcending mere textual interactions. This significant advancement greatly expands the application scenarios of large language models (LLMs).

Addressing this gap is crucial, as the absence of a well-annotated, balanced, and music-centric multi-modal dataset hinders progress in developing models that can effectively understand and generate music based on multi-modal inputs. To overcome this challenge, we introduce a novel dataset comprising 167.69 hours of multi-modal data, including text, images, videos, and music annotations. This dataset is specifically tailored for multi-modal music understanding and generation tasks. Annotated with advanced models such as LLaVA\cite{llava} and Video-LLaVA\cite{videollava}, it offers a rich and diverse set of examples related to music, ensuring the quality and diversity of the data. The creation of this dataset is a pivotal step in advancing the field, as it equips models with the necessary training data to perform effectively across a wide range of music-related tasks.

Large language models are typically composed of a large number of parameters and trained on extensive datasets, endowing them with powerful comprehension and reasoning capabilities. Leveraging these qualities, researchers have utilized LLMs to achieve semantic understanding across various modalities. Examples include engaging in free-form conversations with humans \cite{openai_chatgpt,touvron2023llama}, comprehending audio/video events and performing event-based question answering \cite{tang2023salmonn,huang2023audiogpt,Maaz2023VideoChatGPT,zhao2023learning}, as well as captioning images/3D point cloud data \cite{chen2022visualgpt,li2023blip,xu2023pointllm}. In addition to harnessing the capabilities of LLMs for multi-modal understanding, researchers have also strived to utilize these models to grasp the creative intentions of humans. For instance, they have explored generating images \cite{brade2023promptify}, videos \cite{hong2023cogvideo}, audio \cite{liu2023wavjourney}, or music \cite{copet2023simple} based on textual descriptions, thereby providing valuable assistance in artistic pursuits. Especially for the visual content, some well-performance generative models like Stable Diffusion \cite{stablediffusion,sdvideo} and Sora \cite{sora} can produce high-quality images and videos which even human-eyes cannot distinguish.

Building upon our collected dataset, we propose the \textbf{Mu}lti-modal \textbf{Mu}sic Understanding and Generation using \textbf{LLaMA} (\textbf{MuMu-LLaMA}) framework. This innovative framework leverages novel multi-modal adapters that, in conjunction with encoders such as ViT \cite{dosovitskiy2021image}, ViViT \cite{arnab2021vivit}, and MERT \cite{li2023mert}, effectively capture sequence-level information from various modalities—music, images, and videos—and transform these inputs into rich feature representations. The LLaMA model \cite{touvron2023llama} interprets these features to facilitate both comprehension and generation tasks, tailored to the user's intentions and contextual requirements.

By integrating understanding and generation tasks within the framework of LLMs, we have the potential to significantly enhance the user experience. For example, users can leverage LLMs to summarize videos and generate accompanying audio commentary or suitable background music, thus assisting them in their video creation process. However, research that combines both understanding and generation using LLMs is still limited and in its nascent stage \cite{moon2022multimodal,ge2023planting,huang2023audiogpt,wu2023next,guo2023point,yang2023teal,zhou2024transfusionpredicttokendiffuse,chameleonteam2024chameleonmixedmodalearlyfusionfoundation} especially when it also covers music modality. Among these few existing studies, NExT-GPT \cite{wu2023next} stands out as a significant advancement in the field of multi-modal large language models (MLLMs), excelling in both understanding and generation tasks. Notably, it demonstrates impressive capabilities, including music understanding and generation, image and video question answering, text-to-image and text-to-video generation, as well as audio-driven image and video generation. Despite these advancements, the exploration of music understanding and generation leveraging LLMs remains relatively unexplored. While NExT-GPT exhibits some capabilities in music understanding and generation, its proficiency in music-related tasks is modest due to the absence of specialized training on music datasets. To bridge this gap, we explore the use of LLMs for music understanding and multi-modal music generation.

Our contributions are summarized as follows:
\begingroup
\begin{enumerate}
    \item We present a novel dataset containing 167.69 hours of well-annotated and balanced multi-modal data, annotated with advanced visual models, to support multi-modal music research.
    \item We introduce the MuMu-LLaMA model, a novel data-centric architecture for comprehensive music understanding and multi-modal music generation.
    \item Through rigorous evaluations across four key tasks—music understanding, text-to-music generation, prompt-based music editing, and multi-modal music generation—we demonstrate that MuMu-LLaMA outperforms existing state-of-the-art models, highlighting the potential of combining LLMs with multi-modal inputs for innovative music applications.
\end{enumerate}
\endgroup

\begin{figure*}[t]
    \centering
    \includegraphics[width=\textwidth]{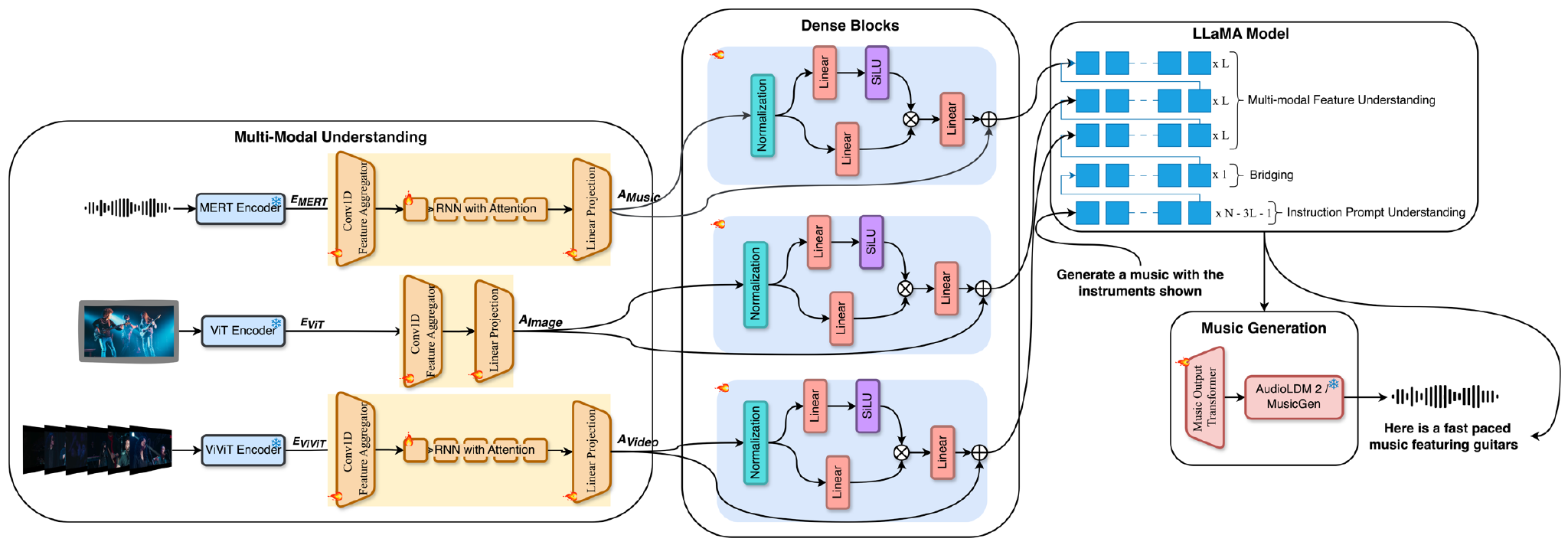}
    \caption{\textbf{Multi-modal Music Understanding and Generation Model (MuMu-LLaMA).} This model framework includes four core components: (1) Pre-trained feature encoders that process inputs from diverse modalities including music, images, and videos. (2) Understanding adapters that integrate these features into a coherent representation suitable for the LLaMA model. (3) The LLaMA model, which contextualizes and interprets the integrated information. (4) An output projection layer that translates the contextual understanding into outputs for the music generation decoder.}
  \label{fig:model}
\end{figure*}
\section{Related Works}
\label{sec:related}

\subsection{Multi-modal Understanding}
The integration of multi-modal data is pivotal in developing AI systems capable of interpreting the complex and heterogeneous information that defines human environments. Research in this domain covers a wide range of tasks, including audio/visual classification \cite{arnab2021vivit}, question answering \cite{lei-etal-2018-tvqa}, captioning \cite{Mei2021act}, tagging \cite{gong2021psla}, event detection \cite{dinkel2021towards}, and summarization \cite{ji2019video}. The emergence of Vision Transformer (ViT) \cite{dosovitskiy2021image} revolutionized the field of computer vision by enabling highly effective visual encoding, leading to the development of methodologies such as ViViT \cite{arnab2021vivit}, which incorporates both temporal and spatial data for enhanced video representation. Similarly, in the domain of music encoding, recent findings have highlighted the superiority of the MERT encoder \cite{li2023mert} in downstream music tagging tasks. In our work, we align with these insights and leverage the MERT encoder to enhance the MuMu-LLaMA framework's ability to comprehend music-related data effectively, thereby ensuring robust performance in multi-modal understanding tasks.

\subsection{Multi-modal Music Generation}
The field of music generation has witnessed substantial advancements, particularly with the adoption of Transformer \cite{vaswani2017attention} and diffusion models \cite{ho2020denoising}, which have significantly elevated the complexity and quality of generative AI outputs. Models such as MusicLM \cite{agostinelli2023musiclm} and MusicGen \cite{copet2023simple} have established new benchmarks in music generation. MusicGen, with its autoregressive Transformer decoder, excels in sequence generation, while AudioLDM 2 \cite{liu2023audioldm2} utilizes a diffusion process to produce high-fidelity audio outputs. Despite these advancements, previous works like Vis2Mus \cite{zhang2022vis2mus} and CMT \cite{di2021video} primarily focus on single-modality music generation, limiting their applicability in diverse contexts. In contrast, our MuMu-LLaMA framework expands these concepts into a comprehensive multi-modal approach, integrating text, image, and video inputs. This approach not only enriches the music generation process but also provides a more holistic understanding and generation of music across various modalities, positioning it as a significant advancement in the field.

\subsection{LLM-assisted Multi-modal Understanding and Generation}
Multi-modal Large Language Models (MLLMs) have emerged as a frontier in AI research, aiming to unify the understanding and generation of diverse data modalities within a single framework. Innovations such as Macaw-LLM \cite{lyu2023macaw} and DreamLLM \cite{dong2023dreamllm} illustrate the potential of integrated multi-modal systems in enhancing user interactions through dynamic content generation. SEED-LLaMA \cite{ge2023making} combines the LLaMA model with diffusion techniques to achieve superior performance in image-related tasks. Similarly, NExT-GPT \cite{wu2023next} introduces a novel approach to manage multi-modal conversations, although it exhibits limitations in handling music-related content due to its restricted music training data. Building on these developments, our contribution, MuMu-LLaMA, specifically addresses the challenges and opportunities in multi-modal music understanding and generation. By enabling the modification of input music based on user prompts and integrating multiple modalities, MuMu-LLaMA significantly enhances AI's role in creative and artistic applications, aligning with the broader objectives of advancing innovative AI technologies with substantial societal and cultural impacts.

\section{MuMu-LLaMA Model Architecture \& Training}

The MuMu-LLaMA model is engineered to harness the synergy of diverse modalities, particularly focusing on music, images, and videos. Figure \ref{fig:model} delineates the architecture which we detail below along with our innovative training methodologies.

\subsection{Multi-modal Feature Encoders}

To minimize training costs while ensuring the multi-modal encoder’s robust capability to handle diverse multi-modal data inputs, MuMu-LLaMA leverages state-of-the-art pre-trained encoders to efficiently extract and integrate complex data across different sensory modalities.

Utilizing the MERT model \cite{li2023mert}, which excels in music tagging \cite{liu2023music}, we encode music features \(\mathbf{X}_\text{Music}\) into embeddings \(\mathbf{E}_\text{MERT}\). The Vision Transformer (ViT) \cite{dosovitskiy2021image} processes images \(\mathbf{X}_\text{Image}\) into feature embeddings \(\mathbf{E}_\text{ViT}\). For video data \(\mathbf{X}_\text{Video}\), the Video Vision Transformer (ViViT) \cite{arnab2021vivit} extracts spatio-temporal tokens, resulting in embeddings \(\mathbf{E}_\text{ViViT}\).


\begin{equation}
\mathbf{E}_\text{MERT} = F_\text{MERT}(\mathbf{X}_\text{Music}) \in \mathbb{R}^{25 \times 1024}
\end{equation}

\begin{equation}
\mathbf{E}_\text{ViT} = F_\text{ViT}(\mathbf{X}_\text{Image}) \in \mathbb{R}^{197 \times 768}
\end{equation}

\begin{equation}
\mathbf{E}_\text{ViViT} = F_\text{ViViT}(\mathbf{X}_\text{Video}) \in \mathbb{R}^{3137 \times 768}
\end{equation}

\noindent
where $F_\text{MERT}(\cdot)$, $F_\text{ViT}(\cdot)$, and $F_\text{ViViT}(\cdot)$ represent feature encoders specifically designed for music, image, and video modalities. The resulting embeddings $\mathbf{E}_\text{MERT}$, $\mathbf{E}_\text{ViT}$, and $\mathbf{E}_\text{ViViT}$ have dimensions of $25 \times 1024$, $197 \times 768$, and $3137 \times 768$.

\subsection{Multi-modal Understanding Adapters}

To integrate diverse modal outputs within the LLaMA framework \cite{touvron2023llama}, we employ multi-modal understanding adapters, which consist of a 1D convolutional layer, linear projection, and a dense network. Since music and video data inherently possess a time dimension and are highly time-dependent, we utilize an RNN coupled with an attention mechanism to project these temporal inputs into a shared 4096-dimensional space. In contrast, image data does not require temporal processing, so we bypass the RNN and attention components for this modality, ensuring efficient and effective cross-modal interaction.

For music and video understanding, the input features \(\mathbf{E}_\text{MERT}\) for music and \(\mathbf{E}_\text{ViViT}\) for video are first encoded using the MERT and ViViT encoders, respectively. These encoded features are then passed through a 1D convolutional layer (Conv1D feature aggregator) to capture local temporal dependencies for better multi-modal alignment. Following this, the features are further processed through an RNN with attention mechanisms to model the sequential nature of the data. The RNN output, $\mathbf{A}_\text{RNN} \in \mathbb{R}^{L \times d}$, where $L$ denotes the sequence length and $d$ is the feature dimension, is used to compute the Query (\(\mathbf{Q}\)), Key (\(\mathbf{K}\)), and Value (\(\mathbf{V}\)) matrices:

\begin{equation}
\mathbf{Q} = \mathbf{A}_\text{RNN}\mathbf{W}^Q, \quad \mathbf{K} = \mathbf{A}_\text{RNN}\mathbf{W}^K, \quad \mathbf{V} = \mathbf{A}_\text{RNN}\mathbf{W}^V
\end{equation}

\noindent
where \(\mathbf{W}^Q \in \mathbb{R}^{d \times d}\), \(\mathbf{W}^K \in \mathbb{R}^{d \times d}\), and \(\mathbf{W}^V \in \mathbb{R}^{d \times d}\) are learnable weight matrices. The attention scores are then calculated as:

\begin{equation}
\text{Attention\_Scores} = \text{Softmax}\left(\frac{\mathbf{Q} \cdot \mathbf{K}^T}{\sqrt{d}}\right)
\end{equation}

Finally, the attention-weighted output is linearly projected to generate the final modality-specific embeddings:

\begin{equation}
\mathbf{A}_\text{Music} = \text{LinearProjection}(\text{Attention\_Scores} \cdot \mathbf{V})
\end{equation}
\begin{equation}
\mathbf{A}_\text{Video} = \text{LinearProjection}(\text{Attention\_Scores} \cdot \mathbf{V})
\end{equation}

In contrast, for image understanding, the input features \(\mathbf{E}_\text{ViT}\) for images are encoded using the ViT encoder. Since image data does not have a temporal dimension, the encoded features skip the RNN and attention layers. Instead, the features are directly passed through the Conv1D Feature Aggregator and then linearly projected to generate the final image-specific embedding:

\begin{equation}
\mathbf{A}_\text{Image} = \text{LinearProjection}(F_\text{Conv1D}(\mathbf{E}_\text{ViT}))
\end{equation}

This method for multi-modal understanding ensures that each modality's unique characteristics are appropriately handled to produce rich, unified representations for subsequent tasks.








\subsection{LLM as a Bridge}


MuMu-LLaMA strategically integrates multi-modal data into LLaMA to enhance context-aware processing. Modality-specific features are introduced every 6 layers within the 32-layer structure, with the last 18 layers divided into three sets of 6 layers (\(L_{18-23}\), \(L_{24-29}\), \(L_{30-35}\)) corresponding to each modality. In each of these sets, hidden states from previous layers are combined with modality-specific features, ensuring that the model effectively processes the input from different modalities.

For each set of layers, the hidden states \(\mathbf{H}_\text{LLaMA}\) from the previous layers are combined with the modality-specific features \(\mathbf{A}_\text{Modality}\) (i.e., \(\mathbf{A}_\text{Video}\), \(\mathbf{A}_\text{Image}\), or \(\mathbf{A}_\text{Music}\)) and a corresponding prefix query \(\mathbf{P}_{\text{query},i}\). The equations governing this integration process are as follows:

\begin{equation}
\mathbf{H}_\text{LLaMA}^{(18-23)} = \text{Layer}_{18-23}\left(\mathbf{H}_\text{LLaMA}^{(12-17)}, \mathbf{A}_\text{Video} + \mathbf{P}_{\text{query},1}\right)
\end{equation}
\begin{equation}
\mathbf{H}_\text{LLaMA}^{(24-29)} = \text{Layer}_{24-29}\left(\mathbf{H}_\text{LLaMA}^{(18-23)}, \mathbf{A}_\text{Image} + \mathbf{P}_{\text{query},2}\right)
\end{equation}
\begin{equation}
\mathbf{H}_\text{LLaMA}^{(30-35)} = \text{Layer}_{30-35}\left(\mathbf{H}_\text{LLaMA}^{(24-29)}, \mathbf{A}_\text{Music} + \mathbf{P}_{\text{query},3}\right)
\end{equation}

Here, \(\mathbf{P}_{\text{query},i} \in \mathbb{R}^{d}\) represents the prefix query for each modality-specific integration step, which provides an additional contextual signal during the multi-modal fusion. The prefix query is crucial for maintaining consistency in the integration process across different layers and modalities. If a modality is unavailable at a particular layer, the model defaults to using only the prefix query for that integration step. The final hidden state \(\mathbf{H}_\text{LLaMA}^{35}\) is then normalized and passed through the LLaMA output layer, completing the multi-modal integration process and ensuring that the combined features are well-prepared for subsequent tasks.

\subsection{Music Understanding and Generation}

Inspired by models like NExT-GPT \cite{wu2023next}, our framework employs discrete audio tokens $[AUD_i]$ ($i \in \{0,1,\cdots,7\}$) to enable dynamic music understanding and generation. This strategy allows for context-sensitive generation of music or text, depending on the task requirements during inference.

\subsection{Training Method}

Considering the computational demands of training from scratch, we utilize the LoRA fine-tuning approach \cite{hu2022lora}, which allows us to effectively adapt MuMu-LLaMA's capabilities while freezing the base encoders and generative models. This not only conserves computational resources but also accelerates the training process. The loss function is designed to optimize the following components depending on the task:

\begin{equation}
    \text{Loss} = 
    \begin{cases} 
        \begin{aligned}
            & L_{CE}(y_{\text{tokens}}, f(y)_{\text{logits}}) \\
            & \quad + \| y_{\text{embeddings}} - g(f(x)_{\text{hidden}}) \|
        \end{aligned}, & \text{if music} \\ \\
        L_{CE}(y_{\text{tokens}}, f(y)_{\text{logits}}), & \text{else}
    \end{cases}
\end{equation}

Here, $L_{CE}$ denotes the cross-entropy loss, crucial for refining text token generation and $g(\cdot)$ denotes the Music Output Transformer. For music generation tasks, we employ mean squared error (MSE) to align the generated embeddings with target music captions, ensuring the high fidelity of generated audio content. An additional regularization term penalizes improper generation of audio tokens, promoting precision in both textual and musical outputs.

\section{Music Oriented Instruction Dataset}
\label{sec:mudataset}

\begin{table*}
  \centering
  \caption{Descriptions of Our Proposed Music Dataset. A/D/R\textsuperscript{*} represents Add/Delete/Replace.}
  \label{tab:dataset}
  \begin{tabular}{cc|ccp{0.5cm}p{0.3cm}|c}
    \hline \hline
    \multicolumn{2}{c|}{Dataset} & \# Audios & Avg. Time (s)/Audio & \multicolumn{2}{p{1.8cm}|}{Total Time (h)} & Data Source \\
    \hline \hline
    \multicolumn{2}{c|}{MUCaps} & 18,515  & 10.00 & \multicolumn{2}{c|}{51.43} & AudioSet \cite{jort2017audioset}\\
    \multicolumn{2}{c|}{MUImage} & 14,520 & 10.00 & \multicolumn{2}{c|}{40.33} & Balanced-AudioSet \cite{jort2017audioset} \\
    \multicolumn{2}{c|}{MUVideo} & 14,504 & 10.00 & \multicolumn{2}{c|}{40.29} & Balanced-AudioSet \cite{jort2017audioset}\\
    \multirow{3}{*}{MUEdit} & Speed & 2384 & 15.11 & 10.01 & \multirow{3}{*}{35.64} & Looperman \cite{Looperman} \\
                            & Pitch & 2369 & 15.20 & 10.00 && Looperman \cite{Looperman}\\
                            & A/D/R\textsuperscript{*} &229 & 245.66 & 15.63 && Slakh \cite{manilow2019cutting}\\
    
  \hline \hline
  \end{tabular}
\end{table*}

Training MLLMs requires extensive data, but there is a shortage of multi-modal datasets focused on music-related tasks. MusicCaps \cite{agostinelli2023musiclm} and MusicQA \cite{liu2023music} are the largest public datasets for music captioning and question answering, but they are insufficient for multi-modal music understanding and generation. For our MuMu-LLaMA model, we need multi-modal instruction datasets for any-to-music generation and extensive datasets such as text-image pairs for alignment training. We use datasets such as Alpaca \cite{alpaca} for instruction following and COCO \cite{lin2014microsoft} for image encoder alignment. Additionally, we collect our own dataset using automated methods inspired by previous works \cite{liu2023music, gong2023listen}, leveraging models like MU-LLaMA \cite{liu2023music}, Mistral-7B-Instruct \cite{mistral}, LLaVA \cite{llava}, and VideoLLaVA \cite{videollava} to perform data annotation.

\begin{figure*}[ht]
\centering

\begin{subfigure}[b]{0.22\textwidth}
    \includegraphics[width=\textwidth]{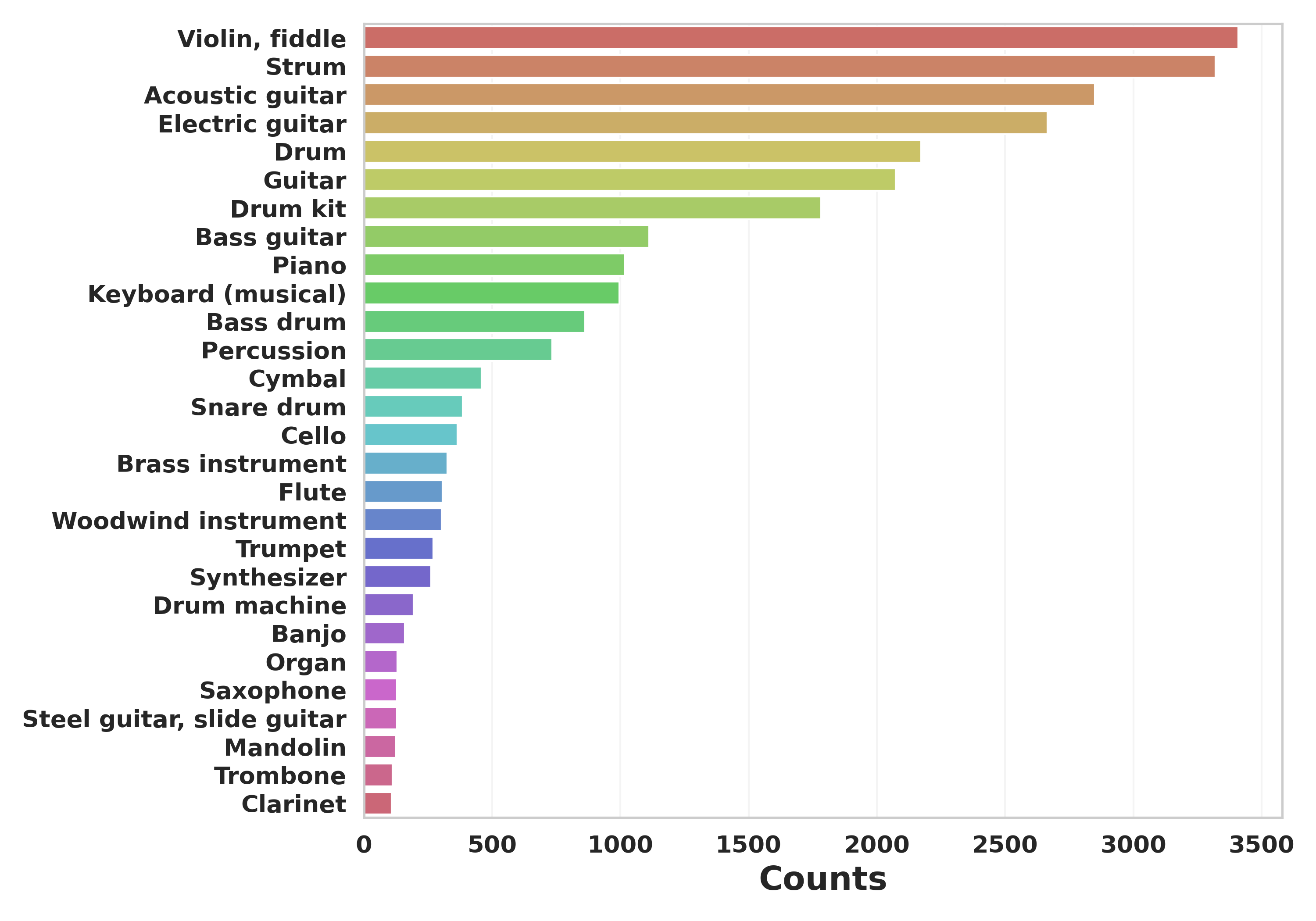}
    \caption{MUCaps}
    \label{fig:sub1}
\end{subfigure}
\hfill
\begin{subfigure}[b]{0.22\textwidth}
    \includegraphics[width=\textwidth]{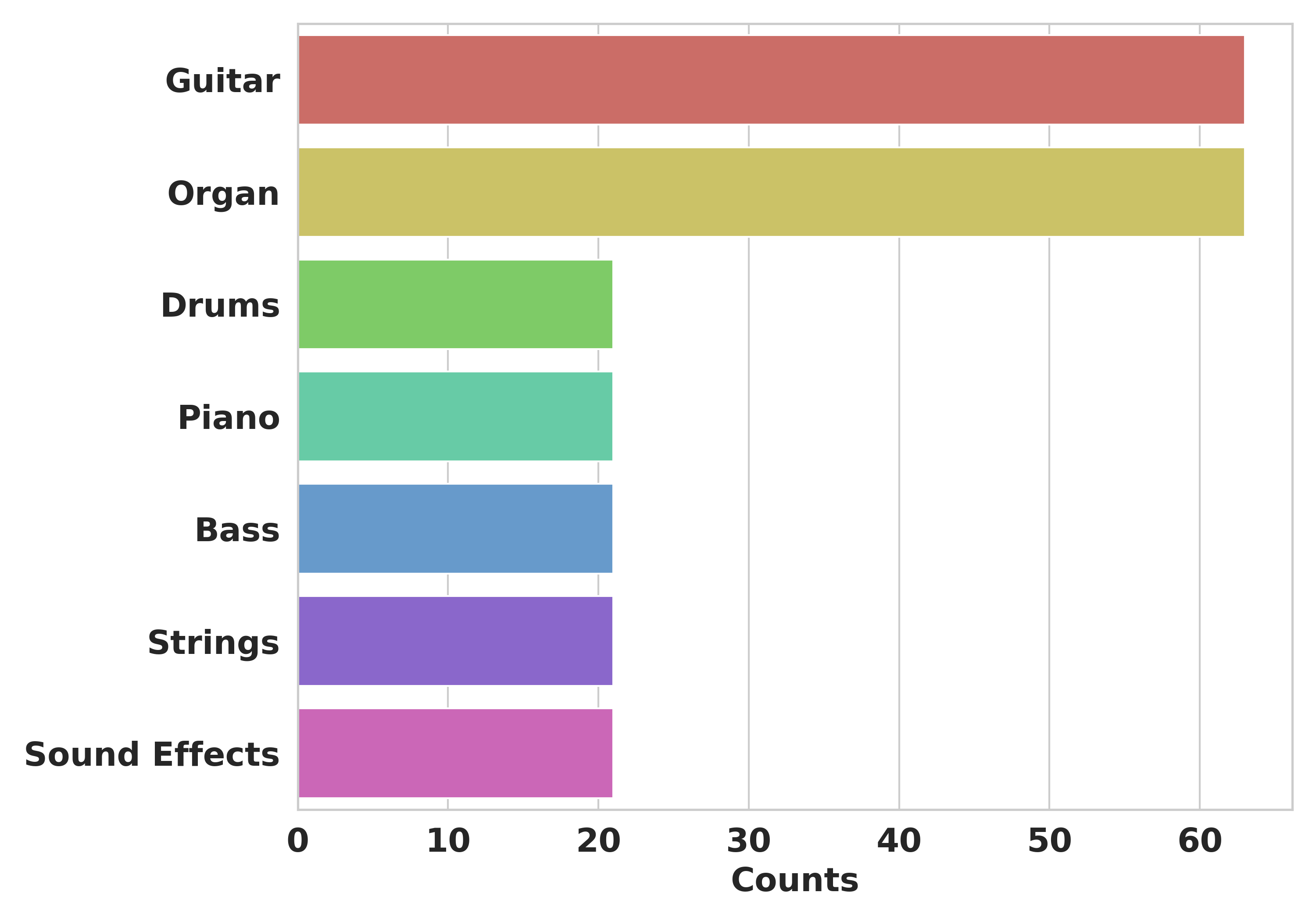}
    \caption{MUEdit - A/D/R}
    \label{fig:sub2}
\end{subfigure}
\hfill
\begin{subfigure}[b]{0.22\textwidth}
    \includegraphics[width=\textwidth]{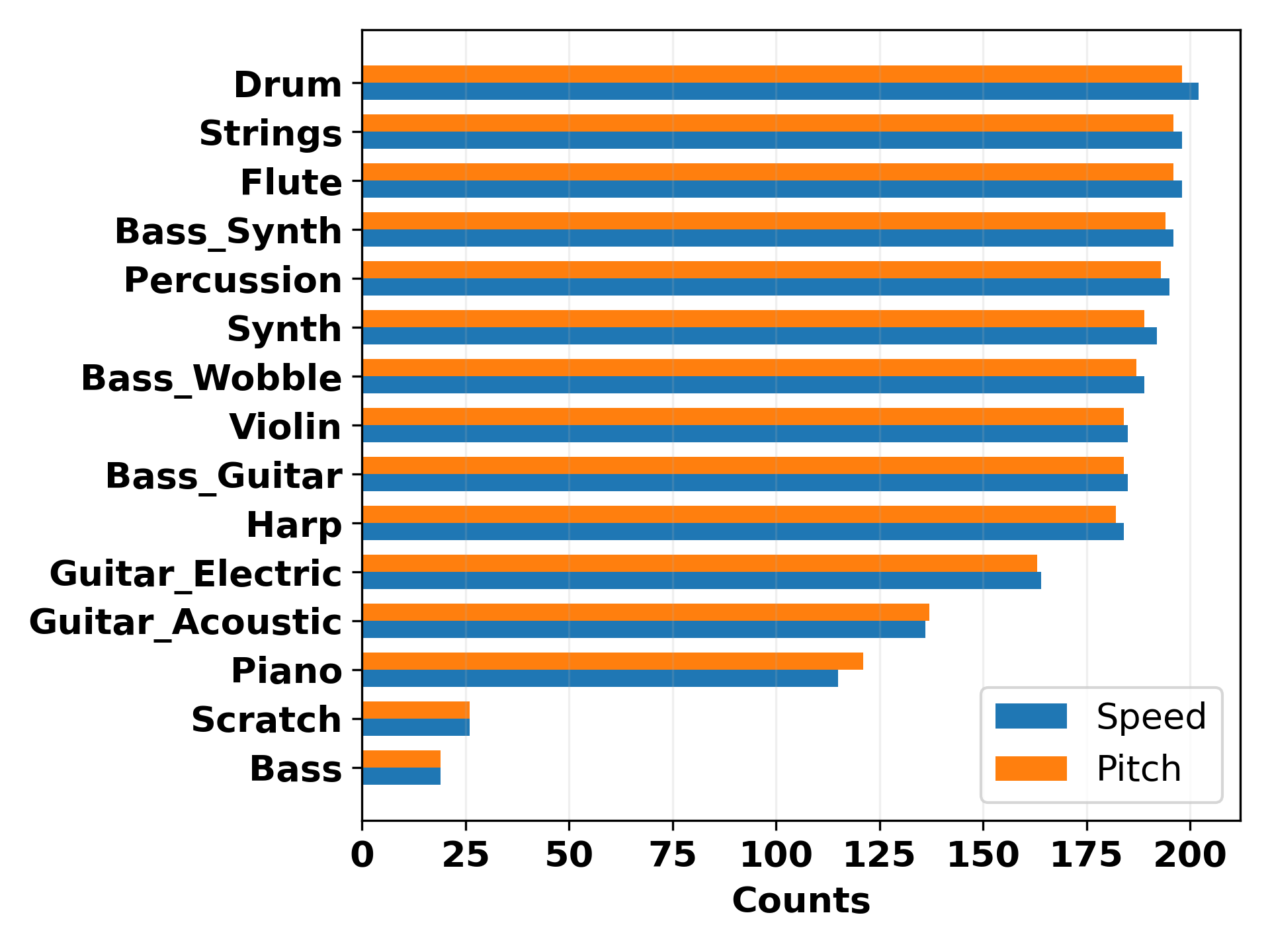}
    \caption{MUEdit - Speed \& Pitch}
    \label{fig:sub3}
\end{subfigure}
\hfill
\begin{subfigure}[b]{0.22\textwidth}
    \includegraphics[width=\textwidth]{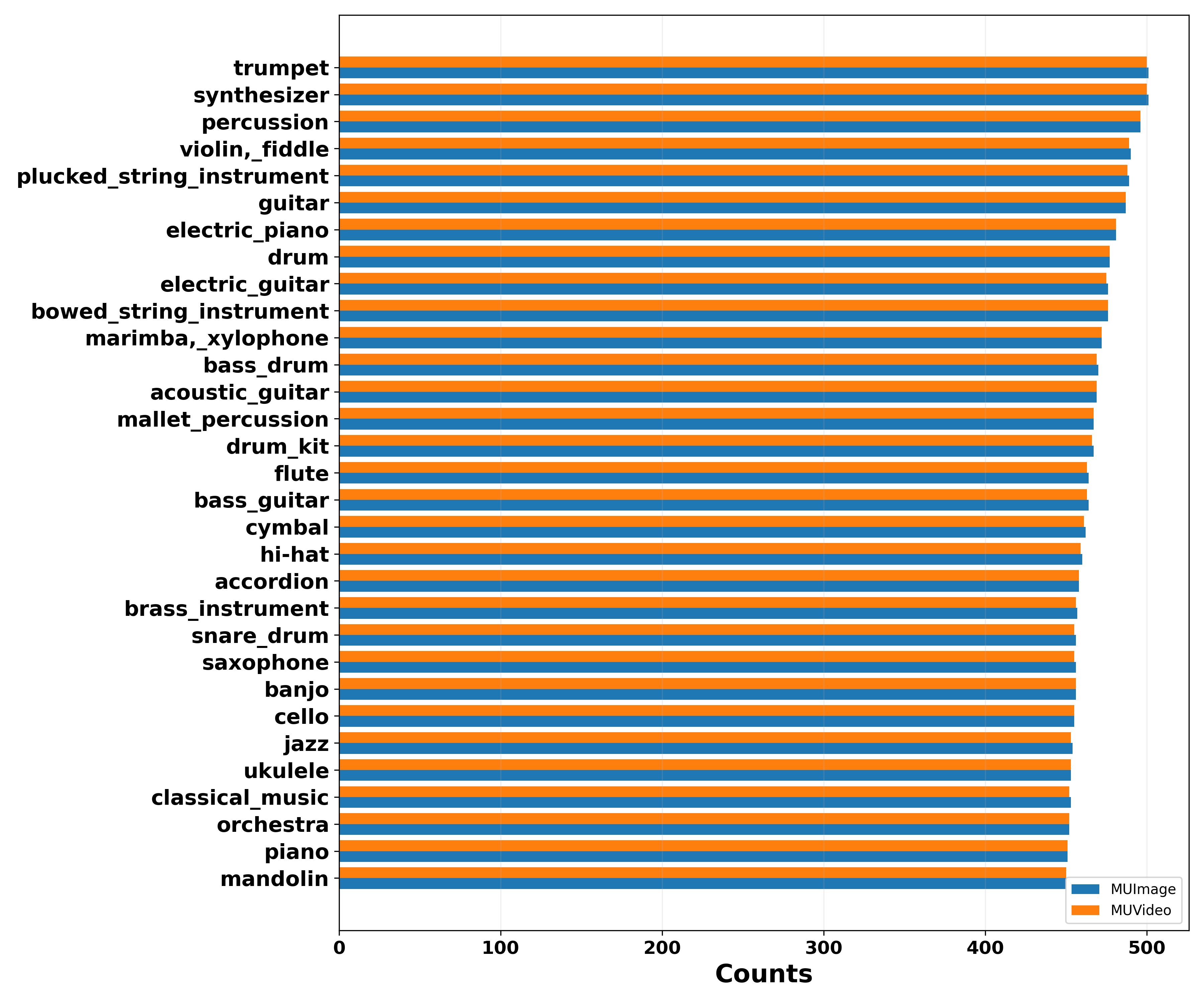}
    \caption{MUImage \& MUVideo}
    \label{fig:sub4}
\end{subfigure}

\caption{\textbf{Distribution of instrument categories in our four curated datasets:} (a) MUCaps reveals a broad diversity of instruments with a long-tail distribution. (b) MUEdit - A/D/R shows a relatively even distribution of add, delete, and replace manipulations across various instruments. (c) MUEdit - Speed \& Pitch demonstrates a consistent distribution of speed and pitch modifications, suggesting balanced attention to tempo and tonal adjustments. (d) MUImage \& MUVideo illustrates a balanced pairing of instruments with corresponding images and videos, ensuring a wide representation within these multi-modal components.}
\label{fig:stats}
\end{figure*}

We create a comprehensive multi-modal dataset with a total of 167.69 hours to enhance MuMu-LLaMA's performance. This fine-grained captioned dataset supports multi-modal understanding and generation. Detailed descriptions are in Table \ref{tab:dataset}. Figure \ref{fig:stats} shows balanced instrument distributions in MUImage and MUVideo datasets. Figures \ref{fig:sub2} and \ref{fig:sub3} indicate a generally balanced distribution in MUEdit, while Figure \ref{fig:sub1} shows a long-tail distribution due to the rarity of certain instruments during data collection and processing.


In the following subsections, we provide a comprehensive overview of the methodologies employed in crafting the datasets used for training the MuMu-LLaMA model.

\subsection{MUCaps Dataset}

We develop the MUCaps dataset, composed of text-music pairs (Table \ref{tab:dataset}), encompassing 51.43 hours of 10-second music files sourced from AudioSet \cite{jort2017audioset} and publicly accessible music websites. The MU-LLaMA model captions the music files with the question: \textit{``Describe the music in detail, including aspects such as instruments used, tempo, and the mood of the song''}. The MUCaps dataset is used for encoder and decoder alignment training.

\subsection{MUEdit Dataset}
To enable music editing in response to prompts, we curated the MUEdit dataset, which includes 35.64 hours of music pairs (Table \ref{tab:dataset}). The dataset generation involves:
\begin{enumerate}[leftmargin=.2in]
    \setlength{\itemsep}{1.5mm}
    \item Use the WSOLA algorithm \cite{wsola} in the sox \cite{Klauer1999SoX} tool to generate speed and pitch-changed music files for origin-to-target pairs, and manipulate individual tracks in the Slakh \cite{manilow2019cutting} dataset for origin-to-Add/Delete/Replace pairs.
    \item Employ the Mistral-7B-Instruct \cite{mistral} model to generate an instruction-response pool with hundreds of templates for each MUEdit subtype, diversifying the dataset and enhancing model robustness.
    \item For each origin-to-X music pair, randomly select instructions and responses from the template pool to construct the final MUEdit dataset.
\end{enumerate}

\subsubsection{Speed}
The WSOLA algorithm modifies music speed without changing pitch to create the Speed split of MUEdit. Supported duration changes are 0.5, 0.7, 1.3, and 1.5 times the original. Instruction-response pairs represent these changes explicitly with numerical values and by degrees of speed variation.

\subsubsection{Pitch}
The Pitch split of MUEdit uses the sox tool's pitch shift feature, allowing pitch changes without altering duration. Permitted pitch variations are $\pm$100 and $\pm$200 cents, representing a semitone and a whole tone. Instructions express pitch variations through verbal expressions, not just numerical values.

\subsubsection{Add/Delete/Replace}
For this branch, we use MIDI data with explicit individual instrument tracks from the Slakh dataset, suitable for tasks like music source separation. In the "Add" sub-split, different tracks from the same MIDI are combined. In the "Delete" sub-split, the mixed track is the input, and individual tracks are the output. In the "Replace" sub-split, tracks from different instruments are selected as input and output.

\subsection{MUImage Dataset}

The MUImage dataset generates fitting music for a given image by pairing music samples from AudioSet with corresponding images. Key steps include:

\begin{enumerate}[leftmargin=.2in]
    \item Use the MU-LLaMA \cite{liu2023music} model to generate music captions for the sampled music files.
    \item Generate captions for the corresponding images using the LLaVA-v1.6-34B model \cite{llava,llavanew} with detailed instructions to describe the image, focusing on instruments and other relevant elements.
    \item Employ the Mistral-7B-Instruct model \cite{mistral} to produce approximately 200 unique instructions that begin with "Generate," such as: "Generate music to match the image."
    \item Generate the model side of the conversation to integrate information from ground-truth tags, music, and image captions, following specific instructions to describe and match the music to the image.
\end{enumerate}

\subsection{MUVideo Dataset}

The MUVideo dataset facilitates video-to-music generation and understanding, sourcing music samples and corresponding videos from the Balanced-AudioSet \cite{jort2017audioset}. Key steps include:

\begin{enumerate}[leftmargin=.2in]
    \item Use the MU-LLaMA model to generate captions for all acquired music files.
    \item Generate captions for the corresponding videos using the VideoLLaVA captioning model \cite{videollava} with detailed instructions to describe the video, focusing on dynamic changes, storyline progression, and visual cues.
    \item Generate both the human and model sides of the conversation using a process similar to the MUImage dataset.
\end{enumerate}

Efforts are made to minimize overlaps among the music files in all datasets. Evaluation sets are established to compare our model's performance with current state-of-the-art (SOTA) models.

\section{Model Evaluation}

We extensively evaluate MuMu-LLaMA on tasks involving music understanding and generation from multi-modal inputs, and compare its performance against other state-of-the-art models. Direct comparison with NExT-GPT \cite{wu2023next} was not feasible due to issues with accessing the required checkpoints. For a fair evaluation, MuMu-LLaMA's hyperparameters were set to a temperature of 0.6, top\_p of 0.8, and a maximum target length of 512 tokens. These hyperparameters were consistently applied across other models in the evaluation, including LLaMA-Adapter \cite{gao2023llamaadapterv2}, MU-LLaMA \cite{liu2023music}, and SALMONN \cite{tang2023salmonn}. Notably, MuMu-LLaMA was paired with the MusicGen decoder, which showed superior performance compared to the AudioLDM 2 decoder. To further understand the contributions of each component within MuMu-LLaMA, an ablation study was conducted.

\subsection{Music Understanding}

MuMu-LLaMA's music understanding capabilities were evaluated using the MTG-eval-QA subset of the MusicQA dataset \cite{liu2023music}, consisting of 4,500 music-related question-answer pairs. The evaluation was conducted against several state-of-the-art (SOTA) models including LTU \cite{gong2023listen}, LLaMA-Adapter \cite{gao2023llamaadapterv2}, SALMONN \cite{tang2023salmonn}, and MU-LLaMA \cite{liu2023music}, the latter being specifically trained on music-related datasets. We employed well-established evaluation metrics such as BLEU (B-U) \cite{papineni2002bleu}, METEOR (M-R) \cite{banerjee2005meteor}, ROUGE$_L$ (R-L) \cite{lin2004rouge}, and BERT-Score (BERT-S) \cite{tianyi2020bertscore} to quantify the model's performance.

Data presented in Table \ref{musicqa_eval} indicate that MuMu-LLaMA outperforms the other models significantly across all metrics. This superior performance is largely attributed to MuMu-LLaMA's advanced music understanding adapter, which incorporates an additional RNN layer and attention mechanism. These components are particularly effective in capturing the temporal information inherent in musical sequences, enabling MuMu-LLaMA to produce more accurate and contextually relevant responses.

\begin{table}[htbp]
\centering
\def\arraystretch{1.1}%
\caption{\textbf{Evaluation of Models on Music Understanding}. The best values of different metrics are made \textbf{bold}.}
\begin{tabular}{c|c|c|c|c}
\hline\hline
\multicolumn{5}{c}{\textbf{Music Understanding}} \\ \hline
Model & \textbf{B-U$\uparrow$} & \textbf{M-R$\uparrow$} & \textbf{R-L$\uparrow$} & \textbf{BERT-S$\uparrow$} \\ \hline\hline
LTU & 0.242 & 0.274 & 0.326 & 0.887 \\ 
LLaMA Adapter & 0.273 & 0.334 & 0.413 & 0.895 \\ 
SALMONN & 0.286 & 0.332 & 0.371 & 0.898 \\
MU-LLaMA & 0.306 & 0.385 & 0.466 & 0.901 \\
\textbf{MuMu-LLaMA} & \textbf{0.341} & \textbf{0.442} & \textbf{0.491} & \textbf{0.908} \\ \hline\hline
\end{tabular}
\label{musicqa_eval}
\end{table}

\begin{table*}[htbp]
\centering
\def\arraystretch{1.2}%
\caption{\textbf{Comparison of Models for Music Generation}. The best values of different metrics are made \textbf{bold}.}
\label{musicgen_eval}
\begin{tabular}{c|ccc|ccc|ccc|ccc}
\hline \hline
\multirow{2}{*}{Model}                    & \multicolumn{3}{c|}{\textbf{Text-to-Music Generation}}                                                                         & \multicolumn{3}{c|}{\textbf{Prompt-based Music Editing}}                                                                    & \multicolumn{3}{c|}{\textbf{Image-to-Music Generation}}                                                                       & \multicolumn{3}{c}{\textbf{Video-to-Music Generation}}                                                                       \\ \cline{2-13} 
                                          & \multicolumn{1}{c|}{\textit{\textbf{FAD$_{vgg}$$\downarrow$}}} & \multicolumn{1}{c|}{\textit{\textbf{KL$\downarrow$}}} & \textit{\textbf{CLAP$_{score}$$\uparrow$}}     & \multicolumn{1}{c|}{\textit{\textbf{FAD$_{vgg}$$\downarrow$}}} & \multicolumn{1}{c|}{\textit{\textbf{KL$\downarrow$}}} & \textit{\textbf{LSD$\downarrow$}} & \multicolumn{1}{c|}{\textit{\textbf{FAD$_{vgg}$$\downarrow$}}} & \multicolumn{1}{c|}{\textit{\textbf{KL$\downarrow$}}} & \textit{\textbf{IB Rank$\uparrow$}} & \multicolumn{1}{c|}{\textit{\textbf{FAD$_{vgg}$$\downarrow$}}} & \multicolumn{1}{c|}{\textit{\textbf{KL$\downarrow$}}} & \textit{\textbf{IB Rank$\uparrow$}} \\ \hline \hline
\textit{CoDi}                             & \multicolumn{1}{c|}{16.201}                & \multicolumn{1}{c|}{6.021}                & 0.143                      & \multicolumn{1}{c|}{N/A}                     & \multicolumn{1}{c|}{N/A}                    & N/A                    & \multicolumn{1}{c|}{10.788}                & \multicolumn{1}{c|}{9.925}                & 0.493                     & \multicolumn{1}{c|}{11.273}                & \multicolumn{1}{c|}{6.267}                & 0.212                     \\ 
\multicolumn{1}{c|}{\textit{AudioLDM 2}} & \multicolumn{1}{c|}{11.619}                & \multicolumn{1}{c|}{4.074}                & \multicolumn{1}{c|}{0.238} & \multicolumn{1}{c|}{N/A}                      & \multicolumn{1}{c|}{N/A}                     & \multicolumn{1}{c|}{N/A} & \multicolumn{1}{c|}{N/A}                      & \multicolumn{1}{c|}{N/A}                     & \multicolumn{1}{c|}{N/A}     & \multicolumn{1}{c|}{N/A}                      & \multicolumn{1}{c|}{N/A}                     & \multicolumn{1}{c}{N/A}     \\ 
\multicolumn{1}{c|}{\textit{MusicGen}}   & \multicolumn{1}{c|}{10.697}                & \multicolumn{1}{c|}{3.909}                & \multicolumn{1}{c|}{0.289} & \multicolumn{1}{c|}{N/A}                      & \multicolumn{1}{c|}{N/A}                     & \multicolumn{1}{c|}{N/A} & \multicolumn{1}{c|}{N/A}                      & \multicolumn{1}{c|}{N/A}                     & \multicolumn{1}{c|}{N/A}     & \multicolumn{1}{c|}{N/A}                      & \multicolumn{1}{c|}{N/A}                     & \multicolumn{1}{c}{N/A}     \\ 
\textit{AUDIT}                            & \multicolumn{1}{c|}{N/A}                      & \multicolumn{1}{c|}{N/A}                     &  \multicolumn{1}{c|}{N/A}                          & \multicolumn{1}{c|}{2.855}                 & \multicolumn{1}{c|}{9.925}                & 0.987                 & \multicolumn{1}{c|}{N/A}                     & \multicolumn{1}{c|}{N/A}                    & N/A                        & \multicolumn{1}{c|}{N/A}                     & \multicolumn{1}{c|}{N/A}                    & N/A                        \\ 
\textit{InstructME}                       & \multicolumn{1}{c|}{N/A}                      & \multicolumn{1}{c|}{N/A}                     & \multicolumn{1}{c|}{N/A}                  & \multicolumn{1}{c|}{2.442}                 & \multicolumn{1}{c|}{6.018}                & 0.846                 & \multicolumn{1}{c|}{N/A}                     & \multicolumn{1}{c|}{N/A}                    & N/A                        & \multicolumn{1}{c|}{N/A}                     & \multicolumn{1}{c|}{N/A}                    & N/A                        \\ 
\textit{CMT}                              & \multicolumn{1}{c|}{N/A}                      & \multicolumn{1}{c|}{N/A}                     & \multicolumn{1}{c|}{N/A}                            & \multicolumn{1}{c|}{N/A}                     & \multicolumn{1}{c|}{N/A}                    & N/A                    & \multicolumn{1}{c|}{N/A}                     & \multicolumn{1}{c|}{N/A}                    & N/A                        & \multicolumn{1}{c|}{9.021}                 & \multicolumn{1}{c|}{5.991}                & 0.629                     \\ 
\textit{MuMu-LLaMA}                       & \multicolumn{1}{c|}{\textbf{9.982}}        & \multicolumn{1}{c|}{\textbf{3.191}}       & \textbf{0.312}             & \multicolumn{1}{c|}{\textbf{1.911}}        & \multicolumn{1}{c|}{\textbf{5.028}}       & \textbf{0.705}        & \multicolumn{1}{c|}{\textbf{6.289}}        & \multicolumn{1}{c|}{\textbf{5.021}}       & \textbf{0.882}            & \multicolumn{1}{c|}{\textbf{7.959}}        & \multicolumn{1}{c|}{\textbf{4.784}}       & \textbf{0.891}            \\ \hline \hline
\end{tabular}
\end{table*}

\subsection{Text-to-Music Generation}

For the task of text-to-music generation, we utilize the MUCaps dataset's 5,000 text-music pairs, comparing MuMu-LLaMA with state-of-the-art models like CoDi \cite{tang2023any}, AudioLDM 2 \cite{liu2023audioldm2}, and MusicGen \cite{copet2023simple}, employing Fr{\'e}chet Audio Distance (FAD) \cite{kilgour2019FrchetAD}, Kullback-Leibler divergence (KL), and CLAP score \cite{laion2023clap} for evaluation. As shown in Table \ref{musicgen_eval}, MuMu-LLaMA demonstrates superior performance, particularly when paired with the MusicGen decoder, which enhances the relevance of the generated music to the input instructions, evidenced by higher CLAP scores. This improvement is largely due to the integration of Large Language Models (LLMs), which enhance the model's comprehension and effective use of input instructions for guiding music generation.

\subsection{Prompt-Based Music Editing}

MuMu-LLaMA stands out as one of the few models supporting music editing through natural language commands, unlike AUDIT \cite{wang2023audit} and InstructME \cite{han2023instructme}, which require specific prompt words like ``Add'' or ``Remove.'' Although Loop Copilot \cite{zhang2023loop} also offers natural language-based editing, it is not open-sourced and thus excluded from our comparison. For AUDIT and InstructME, which are also not open-sourced, we relied on sample outputs available on InstructME's official website for comparison purposes.

To evaluate music editing capabilities, we adopted AUDIT's evaluation metrics, including Fr{\'e}chet Audio Distance (FAD) and Kullback-Leibler divergence (KL), and introduced log spectral distance (LSD) \cite{1162849} for additional assessment. The results in Table \ref{musicgen_eval} show MuMu-LLaMA's superior performance over AUDIT and InstructME, attributed to its use of the LLaMA model for interpreting natural language prompts and the MERT Encoder for understanding source music, significantly enhancing its editing capabilities.

\subsection{Multi-modal Music Generation}

MuMu-LLaMA can generate music from images and videos, a significant feature that sets it apart in the realm of multi-modal music generation. In our experiments, we compare MuMu-LLaMA with CoDi \cite{tang2023any}, an any-to-any generation model capable of both image-to-music (I2M) and video-to-music (V2M) tasks, as well as with CMT \cite{di2021video} specifically for V2M tasks. The evaluation sets for these tasks consist of 2,500 pairs each of image-music and video-music, providing a robust basis for comparison.

To assess the performance of MuMu-LLaMA, we employ traditional metrics such as Fréchet Audio Distance (FAD) and Kullback-Leibler divergence (KL), and introduce ImageBind Ranking (IB Rank) \cite{girdhar2023imagebind}, a novel metric designed to evaluate the alignment between the input modality (image/video) and the generated music. This is achieved by using the ImageBind model to generate embeddings for both the visual input and the corresponding music output, allowing for the calculation of similarity scores that reflect how well the music matches the visual content.

As evidenced by the results in Table \ref{musicgen_eval}, MuMu-LLaMA demonstrates exceptional performance in multi-modal music generation, both in terms of the quality of the music produced and its relevance to the input modality. The model consistently outperforms other state-of-the-art (SOTA) models, highlighting its advanced capability to generate music that is not only high in quality but also closely aligned with the visual content, whether it be from images or videos.

\subsection{Ablation Study}

We evaluated the contributions of the multi-modal understanding adapter's components through an ablation study focused on the dense network and the RNN with attention mechanism for the music understanding task. The results, presented in Table \ref{musicqa_ablation}, provide clear insights into the importance of each component in enhancing MuMu-LLaMA's performance. Specifically, the model variant that excluded both the dense network and the attention RNN showed the lowest performance across all evaluation metrics, indicating the critical role these components play in the architecture.

Adding the dense network to the model led to noticeable improvements in the evaluation metrics, demonstrating its effectiveness in refining the feature representations. Incorporating the RNN component further enhanced the model's performance, with a more significant impact than the dense network alone, suggesting the RNN's crucial role in capturing temporal dependencies in music data. The inclusion of the attention mechanism on top of the RNN provided an additional boost, emphasizing its importance in focusing on the most relevant musical features. Ultimately, the complete MuMu-LLaMA model, utilizing all these components, achieved the best performance, underscoring the synergistic contributions of the dense network, RNN, and attention mechanism in facilitating comprehensive music understanding.

\begin{table}[H]
\centering
\def\arraystretch{1.1}%
\caption{\textbf{Ablation study of our MU-LLaMA model on the music understanding task}. The best values of different metrics are made \textbf{bold}.}
\begin{tabular}{c|c|c|c|c}
\hline\hline
Model & \textbf{B-U$\uparrow$} & \textbf{M-R$\uparrow$} & \textbf{R-L$\uparrow$} & \textbf{BERT-S$\uparrow$} \\ \hline\hline
w/ Projection layer & 0.277 & 0.302 & 0.326 & 0.876 \\ 
w/ Dense Network & 0.303 & 0.354 & 0.401 & 0.880 \\ 
w/ RNN & 0.313 & 0.367 & 0.411 & 0.886 \\
w/ Attn. RNN & 0.336 & 0.375 & 0.439 & 0.894 \\
\textbf{MuMu-LLaMA} & \textbf{0.341} & \textbf{0.442} & \textbf{0.491} & \textbf{0.908} \\ \hline\hline
\end{tabular}
\label{musicqa_ablation}
\end{table}

\subsection{Subjective Evaluation for Music Generation}

To assess our model's music generation capabilities, we conducted a subjective evaluation involving 45 participants. For this evaluation, we designed 13 questions covering three distinct tasks: text-to-music (T2M), image-to-music (I2M), and video-to-music (V2M) generation. Each question presented participants with options generated by different models, and these options were randomly shuffled to eliminate any potential preference bias. This approach ensures that participants' choices were based solely on the quality and relevance of the generated music, rather than any preconceived notions about the models.

In this evaluation, AudioLDM 2 \cite{liu2023audioldm2} and MusicGen \cite{musicgen} were assessed exclusively on the T2M task, while NExT-GPT \cite{wu2023next} and MuMu-LLaMA were evaluated across all three tasks. The CoDi model \cite{codi}, however, was not included in the T2M task due to its limited performance in that area. The results, as presented in Table \ref{subjective_eval}, indicate that our proposed MuMu-LLaMA model consistently received the highest preference among participants across all three music generation tasks. This strong preference highlights MuMu-LLaMA's superior ability to generate music that aligns closely with user inputs across various modalities.

\begin{table}[H]
\centering
\def\arraystretch{1.1}%
\caption{\textbf{Subjective comparison of models for three music generation tasks}. The best values of different metrics are made \textbf{bold}.}
\begin{tabular}{c|c|c|c}
\hline\hline
Model & \textbf{T2M} & \textbf{I2M} & \textbf{V2M} \\ \hline\hline
AudioLDM 2 & 11.6\% & N/A & N/A \\  
MusicGen & 21.3\% & N/A & N/A \\
CoDi & N/A & 5.8\% & 4.4\% \\
NExT-GPT & 8.9\% & 12.9\% & 14.8\% \\
\textbf{MuMu-LLaMA} & \textbf{58.2\%} & \textbf{81.3\%} & \textbf{80.7\%} \\ \hline\hline
\end{tabular}
\label{subjective_eval}
\end{table}

\section{Conclusion}

This paper presents the MuMu-LLaMA model, a novel framework utilizing a large language model (LLM) for integrated music comprehension and multi-modal music generation. Our contributions include not only the development of the model but also a comprehensive methodology for generating specialized datasets to train it. Experimental results demonstrate that MuMu-LLaMA surpasses existing state-of-the-art models in tasks such as music comprehension, music editing, and music generation from text, image, and video inputs. Future work will focus on refining the model's understanding of complex musical nuances and improving the alignment of generated music with diverse multi-modal inputs, further advancing AI's capabilities in creative and cultural applications.

\section{Limitations}

The reliance on the pre-trained MusicGen/AudioLDM 2 model for music generation introduces several challenges that can act as a bottleneck for the overall performance and flexibility of the MuMu-LLaMA model. While these pre-trained models are state-of-the-art in their respective domains, they come with inherent limitations that may affect the MuMu-LLaMA’s ability to generate high-quality, context-aware music.

% ...

\clearpage

\footnotesize
\bibliography{main,sample}

\begin{thebibliography}{77}
\providecommand{\natexlab}[1]{#1}

\bibitem[{Agostinelli et~al.(2023)Agostinelli, Denk, Borsos, Engel et~al.}]{agostinelli2023musiclm}
Agostinelli, A.; Denk, T.~I.; Borsos, Z.; Engel, J.; et~al. 2023.
\newblock MusicLM: Generating Music From Text.
\newblock arXiv:2301.11325.

\bibitem[{Alayrac et~al.(2022)Alayrac, Donahue, Luc, Miech, Barr et~al.}]{alayrac2022flamingo}
Alayrac, J.-B.; Donahue, J.; Luc, P.; Miech, A.; Barr, I.; et~al. 2022.
\newblock {Flamingo: A Visual Language Model for Few-Shot Learning}.
\newblock \emph{Advances in Neural Information Processing Systems}, 35: 23716--23736.

\bibitem[{Arnab et~al.(2021)Arnab, Dehghani, Heigold, Sun, Lucic, and Schmid}]{arnab2021vivit}
Arnab, A.; Dehghani, M.; Heigold, G.; Sun, C.; Lucic, M.; and Schmid, C. 2021.
\newblock {ViViT: A Video Vision Transformer}.
\newblock \emph{2021 IEEE/CVF International Conference on Computer Vision (ICCV)}, 6816--6826.

\bibitem[{Banerjee and Lavie(2005{\natexlab{a}})}]{banerjee2005meteor}
Banerjee, S.; and Lavie, A. 2005{\natexlab{a}}.
\newblock {METEOR: An Automatic Metric for MT Evaluation with Improved Correlation with Human Judgments}.
\newblock In \emph{ACL Workshop}, 65--72.

\bibitem[{Banerjee and Lavie(2005{\natexlab{b}})}]{banerjee-lavie-2005-meteor}
Banerjee, S.; and Lavie, A. 2005{\natexlab{b}}.
\newblock {METEOR: An Automatic Metric for MT Evaluation with Improved Correlation with Human Judgments}.
\newblock In \emph{ACL Workshop}, 65--72.

\bibitem[{Blattmann et~al.(2023)Blattmann, Dockhorn, Kulal, Mendelevitch, Kilian et~al.}]{sdvideo}
Blattmann, A.; Dockhorn, T.; Kulal, S.; Mendelevitch, D.; Kilian, M.; et~al. 2023.
\newblock Stable Video Diffusion: Scaling Latent Video Diffusion Models to Large Datasets.
\newblock arXiv:2311.15127.

\bibitem[{Brade et~al.(2023)Brade, Wang, Sousa, Oore, and Grossman}]{brade2023promptify}
Brade, S.; Wang, B.; Sousa, M.; Oore, S.; and Grossman, T. 2023.
\newblock {Promptify: Text-to-Image Generation through Interactive Prompt Exploration with Large Language Models}.
\newblock In \emph{Proceedings of the 36th Annual ACM Symposium on User Interface Software and Technology}, 1--14.

\bibitem[{Brooks et~al.(2024)Brooks, Peebles, Holmes, DePue et~al.}]{sora}
Brooks, T.; Peebles, B.; Holmes, C.; DePue, W.; et~al. 2024.
\newblock Video generation models as world simulators.

\bibitem[{Chen et~al.(2022)Chen, Guo, Yi, Li, and Elhoseiny}]{chen2022visualgpt}
Chen, J.; Guo, H.; Yi, K.; Li, B.; and Elhoseiny, M. 2022.
\newblock {VisualGPT: Data-efficient Adaptation of Pretrained Language Models for Image Captioning}.
\newblock In \emph{Proceedings of the IEEE/CVF Conference on Computer Vision and Pattern Recognition}, 18030--18040.

\bibitem[{Copet et~al.(2023{\natexlab{a}})Copet, Kreuk, Gat, Remez, Kant, Synnaeve, Adi, and Défossez}]{copet2023simple}
Copet, J.; Kreuk, F.; Gat, I.; Remez, T.; Kant, D.; Synnaeve, G.; Adi, Y.; and Défossez, A. 2023{\natexlab{a}}.
\newblock {Simple and Controllable Music Generation}.
\newblock \emph{arXiv preprint arXiv:2306.05284}.

\bibitem[{Copet et~al.(2023{\natexlab{b}})Copet, Kreuk, Gat, Remez, Kant et~al.}]{musicgen}
Copet, J.; Kreuk, F.; Gat, I.; Remez, T.; Kant, D.; et~al. 2023{\natexlab{b}}.
\newblock Simple and Controllable Music Generation.
\newblock In \emph{Thirty-seventh Conference on Neural Information Processing Systems}.

\bibitem[{Di et~al.(2021)Di, Jiang, Liu, Wang et~al.}]{di2021video}
Di, S.; Jiang, Z.; Liu, S.; Wang, Z.; et~al. 2021.
\newblock {Video Background Music Generation with Controllable Music Transformer}.
\newblock In \emph{Proceedings of the 29th ACM International Conference on Multimedia}, 2037--2045.

\bibitem[{Dinkel, Wu, and Yu(2021)}]{dinkel2021towards}
Dinkel, H.; Wu, M.; and Yu, K. 2021.
\newblock {Towards Duration Robust Weakly Supervised Sound Event Detection}.
\newblock \emph{IEEE/ACM Transactions on Audio, Speech, and Language Processing}, 29: 887--900.

\bibitem[{Dong et~al.(2023)Dong, Han, Peng, Qi, Ge, Yang, Zhao, Sun, Zhou, Wei et~al.}]{dong2023dreamllm}
Dong, R.; Han, C.; Peng, Y.; Qi, Z.; Ge, Z.; Yang, J.; Zhao, L.; Sun, J.; Zhou, H.; Wei, H.; et~al. 2023.
\newblock {DreamLLM: Synergistic Multimodal Comprehension and Creation}.
\newblock \emph{arXiv preprint arXiv:2309.11499}.

\bibitem[{Dosovitskiy et~al.(2021)Dosovitskiy, Beyer, Kolesnikov, Weissenborn, Zhai, Unterthiner, Dehghani, Minderer, Heigold, Gelly, Uszkoreit, and Houlsby}]{dosovitskiy2021image}
Dosovitskiy, A.; Beyer, L.; Kolesnikov, A.; Weissenborn, D.; Zhai, X.; Unterthiner, T.; Dehghani, M.; Minderer, M.; Heigold, G.; Gelly, S.; Uszkoreit, J.; and Houlsby, N. 2021.
\newblock {An Image is Worth 16x16 Words: Transformers for Image Recognition at Scale}.
\newblock In \emph{International Conference on Learning Representations}.

\bibitem[{Gao et~al.(2023)Gao, Han, Zhang, Lin, Geng, Zhou, Zhang, Lu, He, Yue, Li, and Qiao}]{gao2023llamaadapterv2}
Gao, P.; Han, J.; Zhang, R.; Lin, Z.; Geng, S.; Zhou, A.; Zhang, W.; Lu, P.; He, C.; Yue, X.; Li, H.; and Qiao, Y. 2023.
\newblock {LLaMA-Adapter V2: Parameter-Efficient Visual Instruction Model}.
\newblock \emph{arXiv preprint arXiv:2304.15010}.

\bibitem[{Ge et~al.(2023{\natexlab{a}})Ge, Ge, Zeng, Wang, and Shan}]{ge2023planting}
Ge, Y.; Ge, Y.; Zeng, Z.; Wang, X.; and Shan, Y. 2023{\natexlab{a}}.
\newblock {Planting a Seed of Vision in Large Language Model}.
\newblock \emph{arXiv preprint arXiv:2307.08041}.

\bibitem[{Ge et~al.(2023{\natexlab{b}})Ge, Zhao, Zeng, Ge, Li, Wang, and Shan}]{ge2023making}
Ge, Y.; Zhao, S.; Zeng, Z.; Ge, Y.; Li, C.; Wang, X.; and Shan, Y. 2023{\natexlab{b}}.
\newblock {Making LLaMA SEE and Draw with SEED Tokenizer}.
\newblock \emph{arXiv preprint arXiv:2310.01218}.

\bibitem[{Gemmeke et~al.(2017)Gemmeke, Ellis, Freedman, Jansen, Lawrence, Moore, Plakal, and Ritter}]{jort2017audioset}
Gemmeke, J.~F.; Ellis, D. P.~W.; Freedman, D.; Jansen, A.; Lawrence, W.; Moore, R.~C.; Plakal, M.; and Ritter, M. 2017.
\newblock {Audio Set: An Ontology and Human-labeled Dataset for Audio Events}.
\newblock In \emph{Proc. IEEE ICASSP 2017}.

\bibitem[{Girdhar et~al.(2023)Girdhar, El-Nouby, Liu, Singh, Alwala, Joulin, and Misra}]{girdhar2023imagebind}
Girdhar, R.; El-Nouby, A.; Liu, Z.; Singh, M.; Alwala, K.~V.; Joulin, A.; and Misra, I. 2023.
\newblock {ImageBind: One Embedding Space To Bind Them All}.
\newblock In \emph{CVPR}.

\bibitem[{Gong, Chung, and Glass(2021)}]{gong2021psla}
Gong, Y.; Chung, Y.-A.; and Glass, J. 2021.
\newblock {PSLA: Improving Audio Tagging with Pretraining, Sampling, Labeling, and Aggregation}.
\newblock \emph{IEEE/ACM Transactions on Audio, Speech, and Language Processing}, 29: 3292--3306.

\bibitem[{Gong et~al.(2023)Gong, Luo, Liu, Karlinsky, and Glass}]{gong2023listen}
Gong, Y.; Luo, H.; Liu, A.~H.; Karlinsky, L.; and Glass, J. 2023.
\newblock {Listen, Think, and Understand}.
\newblock \emph{arXiv preprint arXiv:2305.10790}.

\bibitem[{Gray and Markel(1976)}]{1162849}
Gray, A.; and Markel, J. 1976.
\newblock Distance measures for speech processing.
\newblock \emph{IEEE Transactions on Acoustics, Speech, and Signal Processing}, 24(5): 380--391.

\bibitem[{Grofit and Lavner(2008)}]{wsola}
Grofit, S.; and Lavner, Y. 2008.
\newblock Time-Scale Modification of Audio Signals Using Enhanced WSOLA With Management of Transients.
\newblock \emph{IEEE Transactions on Audio, Speech, and Language Processing}, 16: 106--115.

\bibitem[{Guo et~al.(2023)Guo, Zhang, Zhu, Tang, Ma, Han, Chen, Gao, Li, Li et~al.}]{guo2023point}
Guo, Z.; Zhang, R.; Zhu, X.; Tang, Y.; Ma, X.; Han, J.; Chen, K.; Gao, P.; Li, X.; Li, H.; et~al. 2023.
\newblock {Point-Bind \& Point-LLM: Aligning Point Cloud with Multi-modality for 3D Understanding, Generation, and Instruction Following}.
\newblock \emph{arXiv preprint arXiv:2309.00615}.

\bibitem[{Han et~al.(2023)Han, Dai, Song, Hao, He, Guo, Chen, Wang, and Qian}]{han2023instructme}
Han, B.; Dai, J.; Song, X.; Hao, W.; He, X.; Guo, D.; Chen, J.; Wang, Y.; and Qian, Y. 2023.
\newblock {InstructME: An Instruction Guided Music Edit And Remix Framework with Latent Diffusion Models}.
\newblock \emph{arXiv preprint arXiv:2308.14360}.

\bibitem[{Ho, Jain, and Abbeel(2020)}]{ho2020denoising}
Ho, J.; Jain, A.; and Abbeel, P. 2020.
\newblock {Denoising Diffusion Probabilistic Models}.
\newblock \emph{Advances in neural information processing systems}, 33: 6840--6851.

\bibitem[{Hong et~al.(2023)Hong, Ding, Zheng, Liu, and Tang}]{hong2023cogvideo}
Hong, W.; Ding, M.; Zheng, W.; Liu, X.; and Tang, J. 2023.
\newblock {CogVideo: Large-scale Pretraining for Text-to-Video Generation via Transformers}.
\newblock In \emph{The Eleventh International Conference on Learning Representations}.

\bibitem[{Hu et~al.(2022)Hu, Shen, Wallis, Allen-Zhu, Li, Wang, Wang, and Chen}]{hu2022lora}
Hu, E.~J.; Shen, Y.; Wallis, P.; Allen-Zhu, Z.; Li, Y.; Wang, S.; Wang, L.; and Chen, W. 2022.
\newblock {LoRA: Low-Rank Adaptation of Large Language Models}.
\newblock In \emph{International Conference on Learning Representations}.

\bibitem[{Huang et~al.(2023)Huang, Li, Yang, Shi, Chang, Ye, Wu, Hong, Huang, Liu et~al.}]{huang2023audiogpt}
Huang, R.; Li, M.; Yang, D.; Shi, J.; Chang, X.; Ye, Z.; Wu, Y.; Hong, Z.; Huang, J.; Liu, J.; et~al. 2023.
\newblock {AudioGPT: Understanding and Generating Speech, Music, Sound, and Talking Head}.
\newblock \emph{arXiv preprint arXiv:2304.12995}.

\bibitem[{Ji et~al.(2019)Ji, Xiong, Pang, and Li}]{ji2019video}
Ji, Z.; Xiong, K.; Pang, Y.; and Li, X. 2019.
\newblock {Video Summarization with Attention-based Encoder-Decoder Networks}.
\newblock \emph{IEEE Transactions on Circuits and Systems for Video Technology}, 30(6): 1709--1717.

\bibitem[{Jiang et~al.(2023)Jiang, Sablayrolles, Mensch, Bamford, Chaplot et~al.}]{mistral}
Jiang, A.~Q.; Sablayrolles, A.; Mensch, A.; Bamford, C.; Chaplot, D.~S.; et~al. 2023.
\newblock Mistral 7B.
\newblock \emph{arXiv preprint arXiv:2310.06825}.

\bibitem[{Kilgour et~al.(2019)Kilgour, Zuluaga, Roblek, and Sharifi}]{kilgour2019FrchetAD}
Kilgour, K.; Zuluaga, M.; Roblek, D.; and Sharifi, M. 2019.
\newblock {Fr{\'e}chet Audio Distance: A Reference-Free Metric for Evaluating Music Enhancement Algorithms}.
\newblock In \emph{Interspeech}.

\bibitem[{Klauer(1999)}]{Klauer1999SoX}
Klauer, U. 1999.
\newblock {SoX - Sound eXchange}.

\bibitem[{Lei et~al.(2018)Lei, Yu, Bansal, and Berg}]{lei-etal-2018-tvqa}
Lei, J.; Yu, L.; Bansal, M.; and Berg, T. 2018.
\newblock {TVQA: Localized, Compositional Video Question Answering}.
\newblock In \emph{Proceedings of the 2018 Conference on Empirical Methods in Natural Language Processing}, 1369--1379. Association for Computational Linguistics.

\bibitem[{Li et~al.(2023{\natexlab{a}})Li, Li, Savarese, and Hoi}]{li2023blip}
Li, J.; Li, D.; Savarese, S.; and Hoi, S. 2023{\natexlab{a}}.
\newblock {BLIP-2: Bootstrapping Language-Image Pre-training with Frozen Image Encoders and Large Language Models}.
\newblock \emph{arXiv preprint arXiv:2301.12597}.

\bibitem[{Li et~al.(2023{\natexlab{b}})Li, Yuan, Zhang, Ma, Chen et~al.}]{li2023mert}
Li, Y.; Yuan, R.; Zhang, G.; Ma, Y.; Chen, X.; et~al. 2023{\natexlab{b}}.
\newblock {MERT: Acoustic Music Understanding Model with Large-Scale Self-supervised Training}.
\newblock \emph{arXiv preprint arXiv:2306.00107}.

\bibitem[{Lin et~al.(2023)Lin, Zhu, Ye, Ning, Jin, and Yuan}]{videollava}
Lin, B.; Zhu, B.; Ye, Y.; Ning, M.; Jin, P.; and Yuan, L. 2023.
\newblock Video-LLaVA: Learning United Visual Representation by Alignment Before Projection.
\newblock \emph{arXiv preprint arXiv:2311.10122}.

\bibitem[{Lin(2004{\natexlab{a}})}]{lin2004rouge}
Lin, C.-Y. 2004{\natexlab{a}}.
\newblock {ROUGE: A Package for Automatic Evaluation of Summaries}.
\newblock In \emph{Text summarization branches out}, 74--81.

\bibitem[{Lin(2004{\natexlab{b}})}]{lin-2004-rouge}
Lin, C.-Y. 2004{\natexlab{b}}.
\newblock {ROUGE: A Package for Automatic Evaluation of Summaries}.
\newblock In \emph{Text summarization branches out}, 74--81.

\bibitem[{Lin et~al.(2014)Lin, Maire, Belongie, Hays et~al.}]{lin2014microsoft}
Lin, T.-Y.; Maire, M.; Belongie, S.; Hays, J.; et~al. 2014.
\newblock {Microsoft COCO: Common Objects in Context}.
\newblock In \emph{Computer Vision--ECCV 2014: 13th European Conference, Zurich, Switzerland, September 6-12, 2014, Proceedings, Part V 13}, 740--755. Springer.

\bibitem[{Liu et~al.(2024)Liu, Li, Li, Li et~al.}]{llavanew}
Liu, H.; Li, C.; Li, Y.; Li, B.; et~al. 2024.
\newblock LLaVA-NeXT: Improved reasoning, OCR, and world knowledge.

\bibitem[{Liu et~al.(2023{\natexlab{a}})Liu, Li, Wu, and Lee}]{llava}
Liu, H.; Li, C.; Wu, Q.; and Lee, Y.~J. 2023{\natexlab{a}}.
\newblock Visual Instruction Tuning.
\newblock arXiv:2304.08485.

\bibitem[{Liu et~al.(2023{\natexlab{b}})Liu, Tian, Yuan, Liu, Mei et~al.}]{liu2023audioldm2}
Liu, H.; Tian, Q.; Yuan, Y.; Liu, X.; Mei, X.; et~al. 2023{\natexlab{b}}.
\newblock {AudioLDM 2: Learning Holistic Audio Generation with Self-supervised Pretraining}.
\newblock \emph{arXiv preprint arXiv:2308.05734}.

\bibitem[{Liu et~al.(2023{\natexlab{c}})Liu, Hussain, Sun, and Shan}]{liu2023music}
Liu, S.; Hussain, A.~S.; Sun, C.; and Shan, Y. 2023{\natexlab{c}}.
\newblock {Music Understanding LLaMA: Advancing Text-to-Music Generation with Question Answering and Captioning}.
\newblock \emph{arXiv preprint arXiv:2308.11276}.

\bibitem[{Liu et~al.(2023{\natexlab{d}})Liu, Zhu, Liu, Yuan, Huang, Liang, Cao, Kong, Plumbley, and Wang}]{liu2023wavjourney}
Liu, X.; Zhu, Z.; Liu, H.; Yuan, Y.; Huang, Q.; Liang, J.; Cao, Y.; Kong, Q.; Plumbley, M.~D.; and Wang, W. 2023{\natexlab{d}}.
\newblock {WavJourney: Compositional Audio Creation with Large Language Models}.
\newblock \emph{arXiv preprint arXiv:2307.14335}.

\bibitem[{Looperman(2000)}]{Looperman}
Looperman. 2000.
\newblock {Looperman - Free Loops, Beats, Samples, Acapellas}.

\bibitem[{Lyu et~al.(2023)Lyu, Wu, Wang, Huang, Liu, Du, Shi, and Tu}]{lyu2023macaw}
Lyu, C.; Wu, M.; Wang, L.; Huang, X.; Liu, B.; Du, Z.; Shi, S.; and Tu, Z. 2023.
\newblock {Macaw-LLM: Multi-Modal Language Modeling with Image, Audio, Video, and Text Integration}.
\newblock \emph{arXiv preprint arXiv:2306.09093}.

\bibitem[{Manilow et~al.(2019)Manilow, Wichern, Seetharaman, and Le~Roux}]{manilow2019cutting}
Manilow, E.; Wichern, G.; Seetharaman, P.; and Le~Roux, J. 2019.
\newblock Cutting Music Source Separation Some {Slakh}: A Dataset to Study the Impact of Training Data Quality and Quantity.
\newblock In \emph{Proc. IEEE Workshop on Applications of Signal Processing to Audio and Acoustics (WASPAA)}. IEEE.

\bibitem[{Mei et~al.(2021)Mei, Liu, Huang, Plumbley, and Wang}]{Mei2021act}
Mei, X.; Liu, X.; Huang, Q.; Plumbley, M.~D.; and Wang, W. 2021.
\newblock {Audio Captioning Transformer}.
\newblock In \emph{Proceedings of the 6th Detection and Classification of Acoustic Scenes and Events 2021 Workshop (DCASE2021)}, 211--215.

\bibitem[{Moon et~al.(2022)Moon, Lee, Shin, Kim, and Choi}]{moon2022multimodal}
Moon, J.~H.; Lee, H.; Shin, W.; Kim, Y.-H.; and Choi, E. 2022.
\newblock {Multi-Modal Understanding and Generation for Medical Images and Text via Vision-Language Pre-Training}.
\newblock \emph{IEEE Journal of Biomedical and Health Informatics}, 26(12): 6070--6080.

\bibitem[{Muhammad~Maaz and Khan(2023)}]{Maaz2023VideoChatGPT}
Muhammad~Maaz, S.~K., Hanoona~Rasheed; and Khan, F. 2023.
\newblock {Video-ChatGPT: Towards Detailed Video Understanding via Large Vision and Language Models}.
\newblock \emph{arXiv preprint arXiv:2306.05424}.

\bibitem[{{OpenAI}(2023)}]{openai_chatgpt}
{OpenAI}. 2023.
\newblock {ChatGPT (Mar 14 version) [Large language model]}.

\bibitem[{Papineni et~al.(2002{\natexlab{a}})Papineni, Roukos, Ward, and Zhu}]{papineni2002bleu}
Papineni, K.; Roukos, S.; Ward, T.; and Zhu, W.-J. 2002{\natexlab{a}}.
\newblock {BLEU: A Method for Automatic Evaluation of Machine Translation}.
\newblock In \emph{Proceedings of the 40th annual meeting of the Association for Computational Linguistics}, 311--318.

\bibitem[{Papineni et~al.(2002{\natexlab{b}})Papineni, Roukos, Ward, and Zhu}]{papineni-etal-2002-bleu}
Papineni, K.; Roukos, S.; Ward, T.; and Zhu, W.-J. 2002{\natexlab{b}}.
\newblock {BLEU: A Method for Automatic Evaluation of Machine Translation}.
\newblock In \emph{ACL}, 311--318.

\bibitem[{Rombach et~al.(2022)Rombach, Blattmann, Lorenz, Esser, and Ommer}]{stablediffusion}
Rombach, R.; Blattmann, A.; Lorenz, D.; Esser, P.; and Ommer, B. 2022.
\newblock High-Resolution Image Synthesis with Latent Diffusion Models.
\newblock arXiv:2112.10752.

\bibitem[{Sun et~al.(2023)Sun, Han, Deng, Wang, Qin, and Gould}]{sun20233d}
Sun, C.; Han, J.; Deng, W.; Wang, X.; Qin, Z.; and Gould, S. 2023.
\newblock {3D-GPT: Procedural 3D Modeling with Large Language Models}.
\newblock \emph{arXiv preprint arXiv:2310.12945}.

\bibitem[{Tang et~al.(2023{\natexlab{a}})Tang, Yu, Sun, Chen, Tan, Li, Lu, Ma, and Zhang}]{tang2023salmonn}
Tang, C.; Yu, W.; Sun, G.; Chen, X.; Tan, T.; Li, W.; Lu, L.; Ma, Z.; and Zhang, C. 2023{\natexlab{a}}.
\newblock {SALMONN: Towards Generic Hearing Abilities for Large Language Models}.
\newblock \emph{arXiv preprint arXiv:2310.13289}.

\bibitem[{Tang et~al.(2023{\natexlab{b}})Tang, Yang, Zhu, Zeng, and Bansal}]{tang2023any}
Tang, Z.; Yang, Z.; Zhu, C.; Zeng, M.; and Bansal, M. 2023{\natexlab{b}}.
\newblock {Any-to-Any Generation via Composable Diffusion}.
\newblock \emph{arXiv preprint arXiv:2305.11846}.

\bibitem[{Tang et~al.(2023{\natexlab{c}})Tang, Yang, Zhu, Zeng, and Bansal}]{codi}
Tang, Z.; Yang, Z.; Zhu, C.; Zeng, M.; and Bansal, M. 2023{\natexlab{c}}.
\newblock Any-to-Any Generation via Composable Diffusion.
\newblock In \emph{Thirty-seventh Conference on Neural Information Processing Systems}.

\bibitem[{Taori et~al.(2023)Taori, Gulrajani, Zhang, Dubois, Li, Guestrin, Liang, and Hashimoto}]{alpaca}
Taori, R.; Gulrajani, I.; Zhang, T.; Dubois, Y.; Li, X.; Guestrin, C.; Liang, P.; and Hashimoto, T.~B. 2023.
\newblock {Stanford Alpaca: An Instruction-following LLaMA model}.
\newblock \url{https://github.com/tatsu-lab/stanford_alpaca}.

\bibitem[{Team(2024)}]{chameleonteam2024chameleonmixedmodalearlyfusionfoundation}
Team, C. 2024.
\newblock Chameleon: Mixed-Modal Early-Fusion Foundation Models.
\newblock arXiv:2405.09818.

\bibitem[{Touvron et~al.(2023)Touvron, Martin, Stone, Albert, Almahairi, Babaei, Bashlykov, Batra, Bhargava, Bhosale et~al.}]{touvron2023llama}
Touvron, H.; Martin, L.; Stone, K.; Albert, P.; Almahairi, A.; Babaei, Y.; Bashlykov, N.; Batra, S.; Bhargava, P.; Bhosale, S.; et~al. 2023.
\newblock {Llama 2: Open Foundation and Fine-tuned Chat Models}.
\newblock \emph{arXiv preprint arXiv:2307.09288}.

\bibitem[{Vaswani et~al.(2017)Vaswani, Shazeer, Parmar, Uszkoreit, Jones, Gomez, Kaiser, and Polosukhin}]{vaswani2017attention}
Vaswani, A.; Shazeer, N.; Parmar, N.; Uszkoreit, J.; Jones, L.; Gomez, A.~N.; Kaiser, {\L}.; and Polosukhin, I. 2017.
\newblock {Attention is All You Need}.
\newblock \emph{Advances in neural information processing systems}, 30.

\bibitem[{Wang et~al.(2023)Wang, Ju, Tan, He, Wu, Bian, and Zhao}]{wang2023audit}
Wang, Y.; Ju, Z.; Tan, X.; He, L.; Wu, Z.; Bian, J.; and Zhao, S. 2023.
\newblock {AUDIT: Audio Editing by Following Instructions with Latent Diffusion Models}.
\newblock \emph{arXiv preprint arXiv:2304.00830}.

\bibitem[{Wu et~al.(2023{\natexlab{a}})Wu, Fei, Qu, Ji, and Chua}]{wu2023next}
Wu, S.; Fei, H.; Qu, L.; Ji, W.; and Chua, T.-S. 2023{\natexlab{a}}.
\newblock {NExT-GPT: Any-to-Any Multimodal LLM}.
\newblock \emph{arXiv preprint arXiv:2309.05519}.

\bibitem[{Wu et~al.(2023{\natexlab{b}})Wu, Chen, Zhang, Hui, Berg-Kirkpatrick, and Dubnov}]{laion2023clap}
Wu, Y.; Chen, K.; Zhang, T.; Hui, Y.; Berg-Kirkpatrick, T.; and Dubnov, S. 2023{\natexlab{b}}.
\newblock {Large-scale Contrastive Language-Audio Pretraining with Feature Fusion and Keyword-to-Caption Augmentation}.
\newblock In \emph{IEEE International Conference on Acoustics, Speech and Signal Processing, ICASSP}.

\bibitem[{Xu et~al.(2023{\natexlab{a}})Xu, Liu, Khan, Fan, and Wu}]{pmlr-v209-xu23a}
Xu, L.; Liu, B.; Khan, A.~H.; Fan, L.; and Wu, X.-M. 2023{\natexlab{a}}.
\newblock {Multi-modal Pre-training for Medical Vision-language Understanding and Generation: An Empirical Study with A New Benchmark}.
\newblock In \emph{Proceedings of the Conference on Health, Inference, and Learning}, volume 209 of \emph{Proceedings of Machine Learning Research}, 117--132.

\bibitem[{Xu et~al.(2023{\natexlab{b}})Xu, Wang, Wang, Chen, Pang, and Lin}]{xu2023pointllm}
Xu, R.; Wang, X.; Wang, T.; Chen, Y.; Pang, J.; and Lin, D. 2023{\natexlab{b}}.
\newblock {PointLLM: Empowering Large Language Models to Understand Point Clouds}.
\newblock \emph{arXiv preprint arXiv:2308.16911}.

\bibitem[{Yang et~al.(2023)Yang, Zhang, Meng, and Zhou}]{yang2023teal}
Yang, Z.; Zhang, Y.; Meng, F.; and Zhou, J. 2023.
\newblock {TEAL: Tokenize and Embed ALL for Multi-modal Large Language Models}.
\newblock \emph{arXiv preprint arXiv:2311.04589}.

\bibitem[{Yin et~al.(2023)Yin, Fu, Zhao, Li et~al.}]{yin2023survey}
Yin, S.; Fu, C.; Zhao, S.; Li, K.; et~al. 2023.
\newblock {A Survey on Multimodal Large Language Models}.
\newblock \emph{arXiv preprint arXiv:2306.13549}.

\bibitem[{Zhang et~al.(2022)Zhang, Zhang, Shao, Shan, and Xia}]{zhang2022vis2mus}
Zhang, R.; Zhang, Y.; Shao, K.; Shan, Y.; and Xia, G. 2022.
\newblock {Vis2Mus: Exploring Multimodal Representation Mapping for Controllable Music Generation}.
\newblock \emph{arXiv preprint arXiv:2211.05543}.

\bibitem[{Zhang* et~al.(2020{\natexlab{a}})Zhang*, Kishore*, Wu*, Weinberger, and Artzi}]{tianyi2020bertscore}
Zhang*, T.; Kishore*, V.; Wu*, F.; Weinberger, K.~Q.; and Artzi, Y. 2020{\natexlab{a}}.
\newblock {BERTScore: Evaluating Text Generation with BERT}.
\newblock In \emph{ICLR}.

\bibitem[{Zhang* et~al.(2020{\natexlab{b}})Zhang*, Kishore*, Wu*, Weinberger, and Artzi}]{bert-score}
Zhang*, T.; Kishore*, V.; Wu*, F.; Weinberger, K.~Q.; and Artzi, Y. 2020{\natexlab{b}}.
\newblock {BERTScore: Evaluating Text Generation with BERT}.
\newblock In \emph{ICLR}.

\bibitem[{Zhang et~al.(2023)Zhang, Maezawa, Xia, Yamamoto, and Dixon}]{zhang2023loop}
Zhang, Y.; Maezawa, A.; Xia, G.; Yamamoto, K.; and Dixon, S. 2023.
\newblock {Loop Copilot: Conducting AI Ensembles for Music Generation and Iterative Editing}.
\newblock \emph{arXiv preprint arXiv:2310.12404}.

\bibitem[{Zhao et~al.(2023)Zhao, Misra, Kr{\"a}henb{\"u}hl, and Girdhar}]{zhao2023learning}
Zhao, Y.; Misra, I.; Kr{\"a}henb{\"u}hl, P.; and Girdhar, R. 2023.
\newblock {Learning Video Representations from Large Language Models}.
\newblock In \emph{Proceedings of the IEEE/CVF Conference on Computer Vision and Pattern Recognition}, 6586--6597.

\bibitem[{Zhou et~al.(2024)Zhou, Yu, Babu, Tirumala, Yasunaga, Shamis, Kahn, Ma, Zettlemoyer, and Levy}]{zhou2024transfusionpredicttokendiffuse}
Zhou, C.; Yu, L.; Babu, A.; Tirumala, K.; Yasunaga, M.; Shamis, L.; Kahn, J.; Ma, X.; Zettlemoyer, L.; and Levy, O. 2024.
\newblock Transfusion: Predict the Next Token and Diffuse Images with One Multi-Modal Model.
\newblock arXiv:2408.11039.

\end{thebibliography}

\clearpage

\section{Appendix}
In the appendix, comprehensive information is provided concerning the model's training dataset and training methodology, encompassing explicit insights into the utilized training approach and the corresponding model hyperparameters. Additionally, a thorough exposition is given regarding the composition of the evaluation sets employed in our study, accompanied by a delineation of the evaluation methodology and metrics applied to assess the performance of our model. To elucidate the diverse capabilities of our model, illustrative demo examples are also included.

\section{Music Oriented Dataset Information}

\begin{figure*}[!t]
    \centering
    \includegraphics[width=\textwidth]{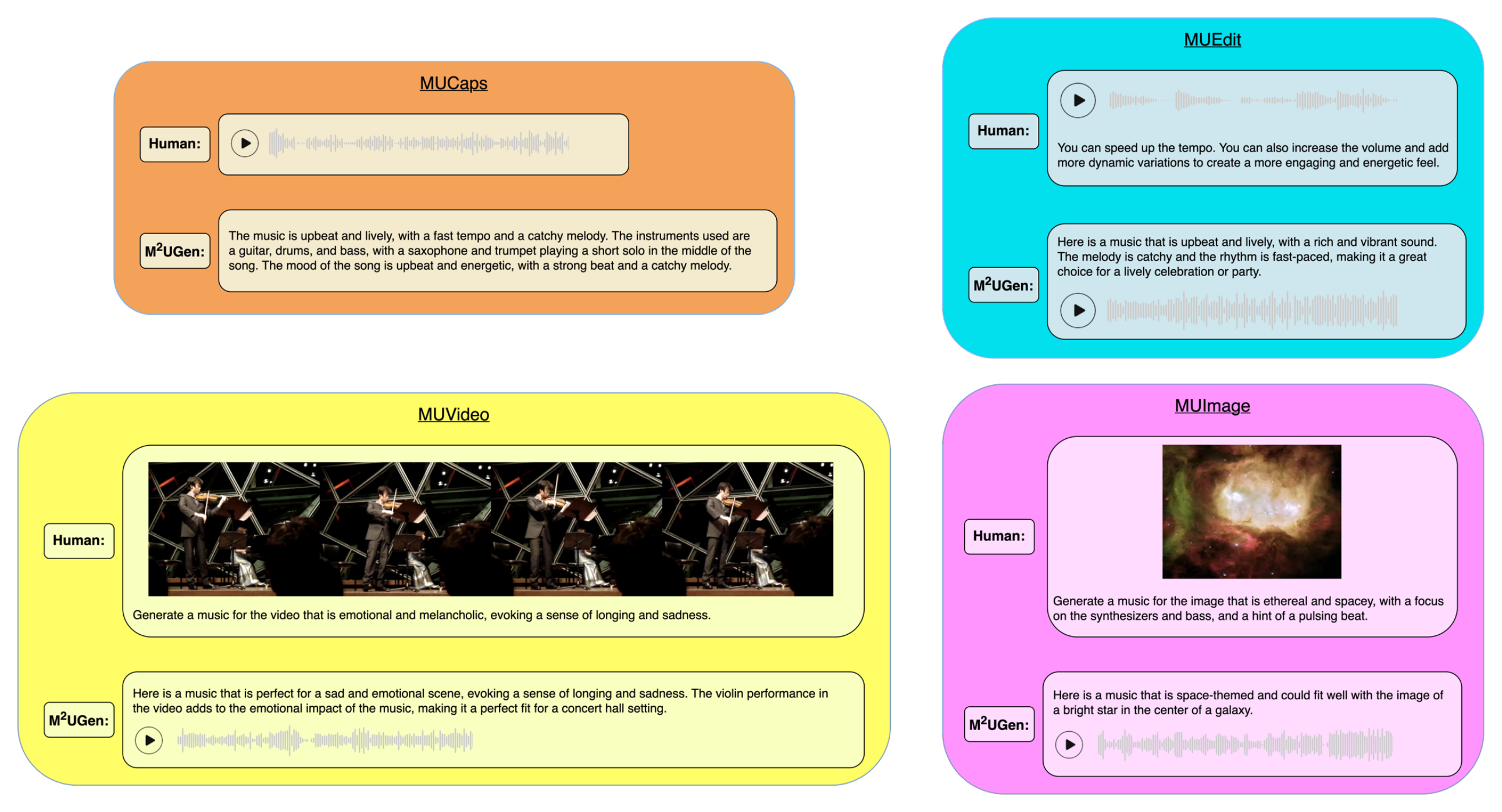}
    \caption{\textbf{Music Oriented Dataset.} Examples from the MUCaps, MUEdit, MUImage and MUVideo datasets used to train the MuMu-LLaMA model.}
    \label{fig:dataset}
\end{figure*}

We generate 4 different datasets to train the MuMu-LLaMA model: MUCaps, MUImage, MUVideo and MUEdit datasets. An example of each from the 4 datasets are shown in Figure \ref{fig:dataset}.

\section{Model Training}

In this section, we detail the training strategy for the MuMu-LLaMA model along with parameters used for training.

\subsection{Model Training Strategy}

\begin{figure*}[!t]
    \centering
    \includegraphics[width=\textwidth]{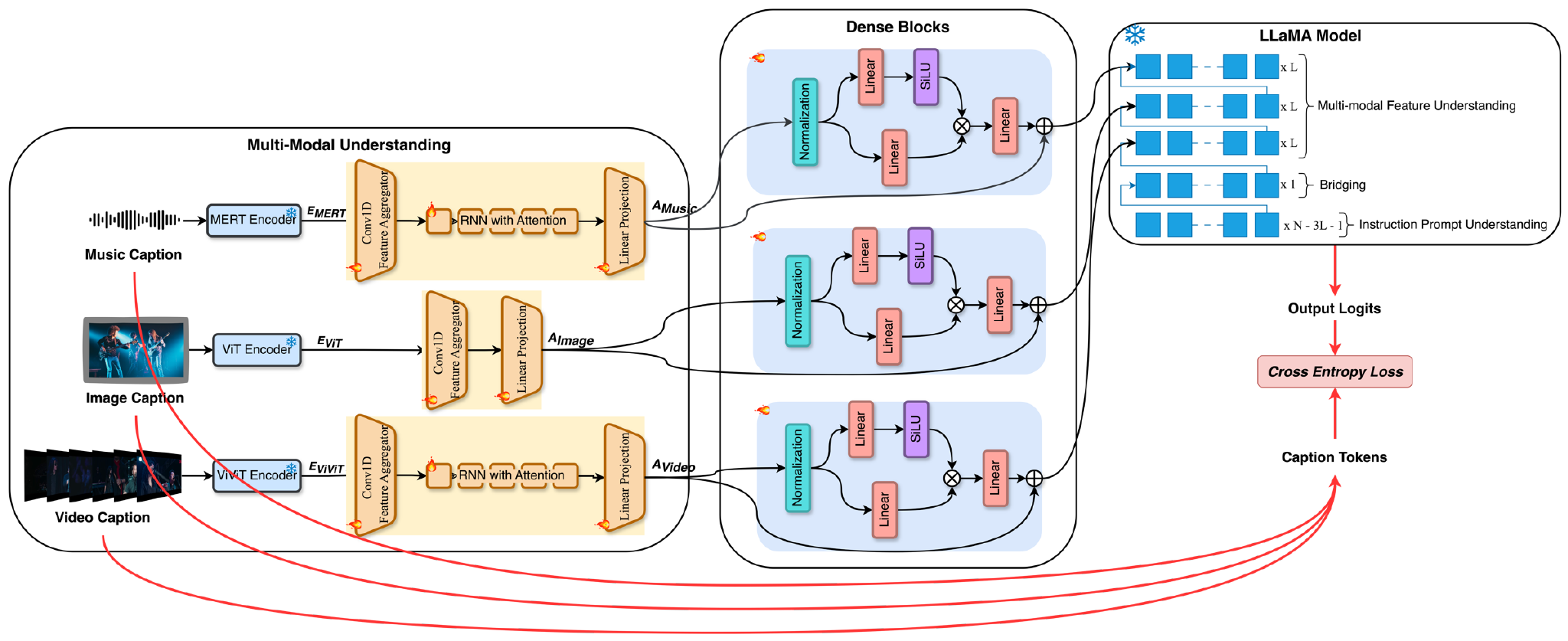}
    \caption{\textbf{Training Stage 1}: The Multi-modal Understanding Adapters are trained to integrate multi-modal features into the different layers of the LLaMA model.}
    \label{fig:stage1}
\end{figure*}

The MuMu-LLaMA model adopts the adapter training strategy, implementing a three-step training regimen. In the first phase, all parameters, with the exception of those associated with the Multi-modal Understanding Adapters, undergo freezing. The training dataset is configured to incorporate the MUCaps dataset for music understanding, the COCO dataset for image comprehension, and the captions sourced from the MUVideo dataset for video understanding. During this training stage, the Cross Entropy Loss function is applied to compute the disparity between the caption generated by the LLaMA model and the target caption corresponding to the input modality. This process is illustrated in Figure \ref{fig:stage1}.

In the second training stage, the output projector is trained to generate conditional embeddings using input captions processed by the LLaMA model. The LLaMA model produces specialized audio tokens, denoted as [AUD$i$] where $i \in \{1, 2, \ldots, K\}$ (with $K$ as a hyperparameter representing the number of special audio tokens added to the LLaMA model's vocabulary) when processing input captions. The special audio tokens serve as signaling indicators, aiding the model in determining whether to generate text+music or solely text. In training, these audio tokens are added to the end of the text output in datasets requiring music output. During inference, if the MuMu-LLaMA model generates audio tokens, downstream music decoders (MusicGen/AudioLDM 2) will perform music generation, otherwise, solely text will be produced.

The hidden embeddings corresponding to these audio tokens from the last layer of the LLaMA model is then input into the output projection layer, generating the conditional embedding for the Music Generation model. The MUCaps dataset is utilized to train this stage, with captions serving as inputs to the model and the target output tokens set as the special audio tokens.

\begin{figure*}[!t]
    \centering
    \includegraphics[width=\textwidth]{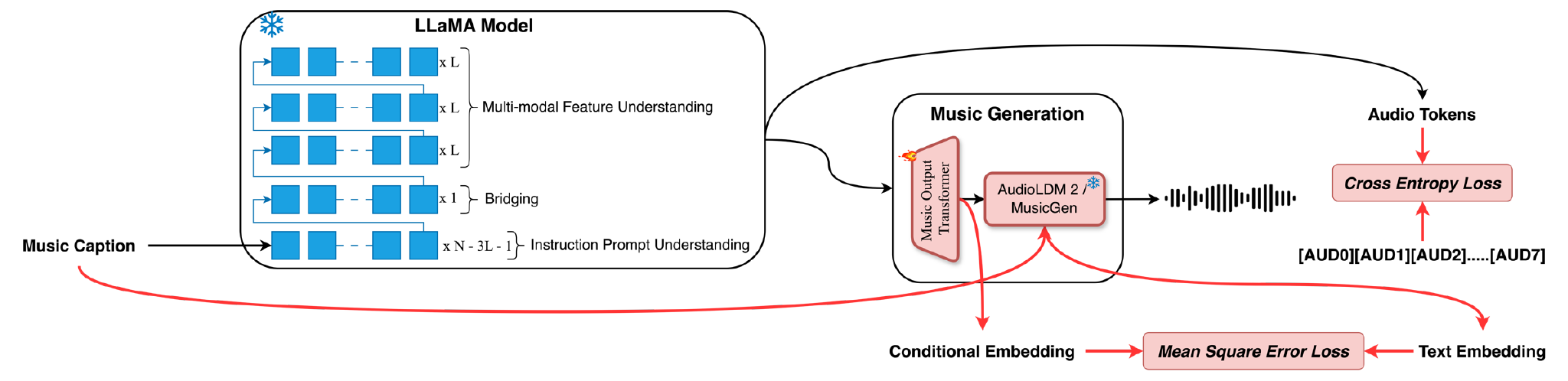}
    \caption{\textbf{Training Stage 2}: The Output Projection Layer is trained to generate the conditioning embedding for the MusicGen/AudioLDM 2 model.}
    \label{fig:stage2}
\end{figure*}

Assuming a total of $N$ tokens generated by the LLaMA model, where [AUD$i$] with $i \in \{0, 2, \ldots, 7\}$ constitutes the last $8$ tokens. The hidden embeddings size is $(1, N, 4096)$, and the last 8 tokens are extracted along dimension $-1$, resulting in an input embedding size of the Output Projection layer as $(1, 8, 4096)$. The output size from the projection layer varies based on the Music Generation model: for AudioLDM2, it is $(1, 512)$, and for MusicGen, it is $(512, 768)$.

\vspace{2mm}

In the final training stage, the LoRA training strategy is employed to train the LLaMA model, concurrently fine-tuning the Multi-modal Understanding Adapter and Output Projection layer. This stage utilizes datasets including Alpaca, MusicQA, MUImage, MUVideo, and MUEdit. To signal the MuMu-LLaMA model to generate both music and text, the output text in MUImage, MUVideo, and MUEdit datasets is extended with the special audio tokens.

\begin{figure*}[!t]
    \centering
    \includegraphics[width=\textwidth]{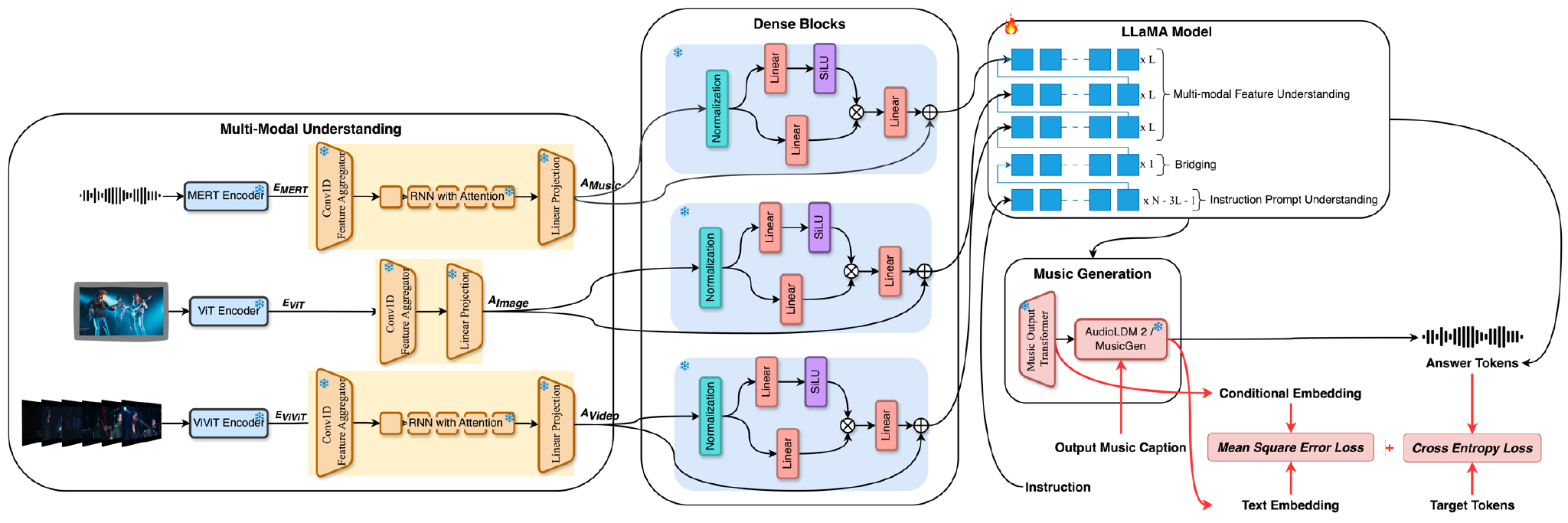}
    \caption{\textbf{Training Stage 3}: The Multi-modal Understanding Adapter and Output Projection Layer are fine-tuned while the LoRA-enabled LLaMA model is trained in this stage.}
    \label{fig:stage3}
\end{figure*}

\subsection{Reasoning for Training Strategy}

Using the initial training stage, the model undergoes training with the objective of comprehending diverse modalities by leveraging extensive captioning datasets for music, image, and video. The subsequent training stage focuses on refining the LLaMA model's capability to condition music generation based on input captions.

These dual training stages equip the model with the ability to grasp various modalities and generate distinct music conditions based on input captions. This proficiency significantly contributes to the final training stage. In this ultimate phase, the model leverages the trained Multi-modal Understanding Adapters and Output Projection layers to bridge the gap between them, honing the LLaMA model's skills through the utilization of multi-modal music generation and music understanding datasets.

\subsection{Model Training Parameters}

We conduct training for the three stages of our model, employing 5, 5, and 2 epochs, respectively. The training process incorporates the following hyper-parameters: $N=32$, $L=6$, $\text{number of Audio Tokens}=8$, and $lr=10^{-4}$. This choice of hyper-parameters, coupled with our training strategy, allows for the effective use of a reduced number of epochs and a smaller dataset during the final stage of model training.

\section{Model Evaluation}

In this section, we elaborate on the datasets employed to assess the various capabilities of the MuMu-LLaMA model, followed by a discussion of the evaluation metrics utilized.

\subsection{Evaluation Datasets}

For each of the training datasets generated—MUCaps, MUImage, MUVideo, and MUEdit—we create a corresponding evaluation set. The methodology employed for generating the evaluation dataset mirrors that of the training dataset generation. Detailed statistics for the evaluation datasets are provided in Table \ref{eval_stats}. However, for the MUEdit dataset, the evaluation set could not be utilized for model evaluation due to the unavailability of code bases and trained checkpoints for InstructME\cite{han2023instructme} and AUDIT\cite{wang2023audit}. Consequently, we resort to utilizing samples from InstructME's demo website, which includes samples from both AUDIT and InstructME, to assess our model's performance. For evaluating the MuMu-LLaMA's music understanding capabilities we utilize the evaluation split of the MusicQA dataset.

\begin{table}[!t]
\centering
\def\arraystretch{1.2}%
\caption{\textbf{Evaluation Dataset Statistics}. The number of instructions in the evaluation dataset and total hours of music files in the dataset}
\begin{tabular}{c|c|c}
\hline \hline
\textbf{Dataset} & \textbf{Number of Instructions} & \textbf{Hours of Music} \\ \hline \hline
\textbf{MUCaps Eval}  & 4000                           & 265.35                 \\ \hline
\textbf{MUImage Eval} & 2500                            & 6.94                   \\ \hline
\textbf{MUVideo Eval} & 2500                           & 6.94                   \\ \hline
\textbf{MUEdit Eval}  & 2000                           & 5.55                   \\ \hline \hline
\end{tabular}
\label{eval_stats}
\end{table}

\subsection{Evaluation Metrics}

To assess music question answering, we adopt the metrics employed in \cite{liu2023music}, namely BLEU (B-U) \cite{papineni-etal-2002-bleu}, METEOR (M-R) \cite{banerjee-lavie-2005-meteor}, ROUGE$_L$ (R-L) \cite{lin-2004-rouge}, and BERT-Score (BERT-S) \cite{bert-score}. These metrics are widely used for evaluating text generation. For all music generation tasks, we employ Fr{\'e}chet Audio Distance (FAD) \cite{kilgour2019FrchetAD} and Kullback-Leibler divergence (KL), as these metrics are commonly utilized to assess the quality of generated audio. In addition to these general metrics, task-specific metrics are applied for each of the music generation tasks, namely Text-to-Music, Image-to-Music, Video-to-Music, and Music Editing.

In the context of Text-to-Music, we employ the CLAP\cite{laion2023clap} score, calculated by determining the cosine similarity between the CLAP embedding for the generated music and the text input:
\[
    CLAPScore(M, T) = \max(100 \times \cos(E_M, E_T), 0)
\]
Here, $M$ represents the generated music, $T$ denotes the text input, and $E_M$, $E_T$ represent the CLAP embeddings for the music and text, respectively.

For the Music Editing task, we leverage the Log Spectral Distance (LSD) to assess the disparity between the generated music and the target music. This metric facilitates the evaluation of whether the frequencies in the generated music, post-editing, align with those in the target music.

For Image-to-Music and Video-to-Music tasks, we introduce the ImageBind\cite{girdhar2023imagebind} Ranking (IB Rank), akin to the CLAP score, to quantify the alignment between the input modality and the generated music. Considering $N$ distinct models producing $N$ music files, we generate ImageBind embeddings for the music files, denoted as $E_{M1}, E_{M2}, \ldots, E_{MN}$, as well as the ImageBind embedding for the input modality, denoted as $E_{I/V}$. The embeddings for the music files are ranked based on their cosine similarity to $E_{I/V}$. Subsequently, after ranking all the samples in the evaluation set, the ImageBind Ranking is computed by calculating the ranking score using the individual rankings.

Using these evaluation metrics, we are able to evaluate the MuMu-LLaMA model against other state-of-the-art models for the different tasks.

\section{Model Demonstration}

In this section, we present screenshots of the MuMu-LLaMA model demo, illustrating various capabilities of the model.

\begin{figure*}[!t]
    \centering
    \includegraphics[width=\textwidth]{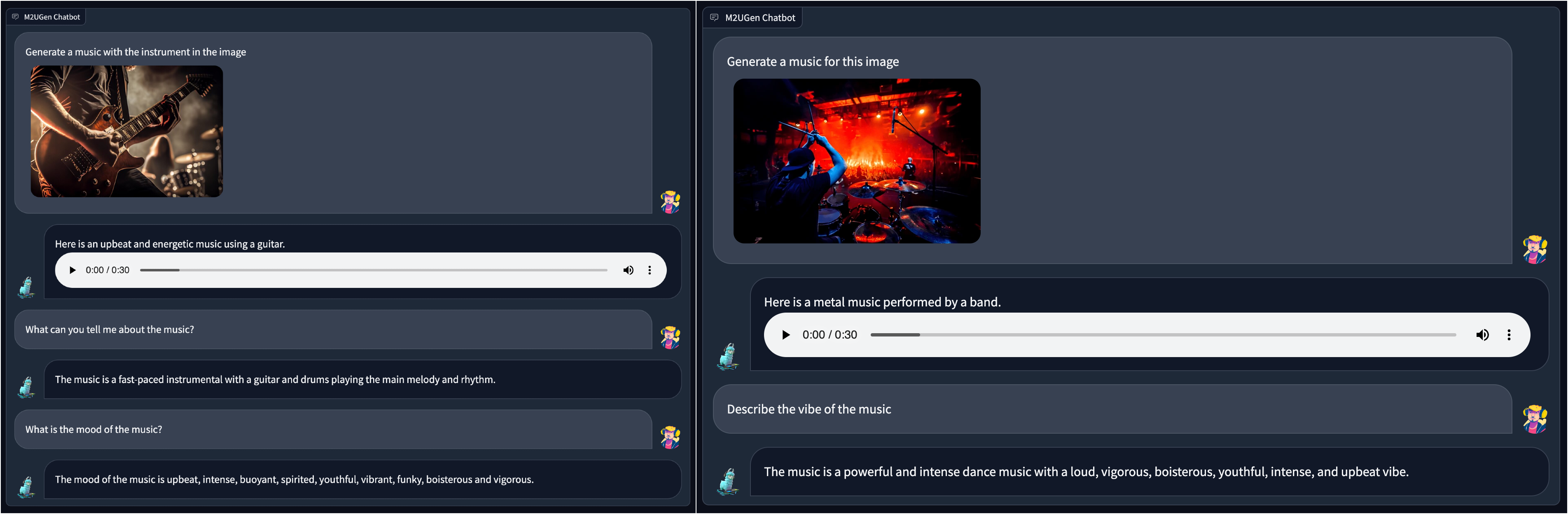}
    \caption{\textbf{Image-To-Music Generation and Understanding:} The MuMu-LLaMA model is capable of generating music for images and also answering questions regarding the generated music.}
    \label{fig:I2M+MU}
\end{figure*}

\begin{figure*}[!t]
    \centering
    \includegraphics[width=\textwidth]{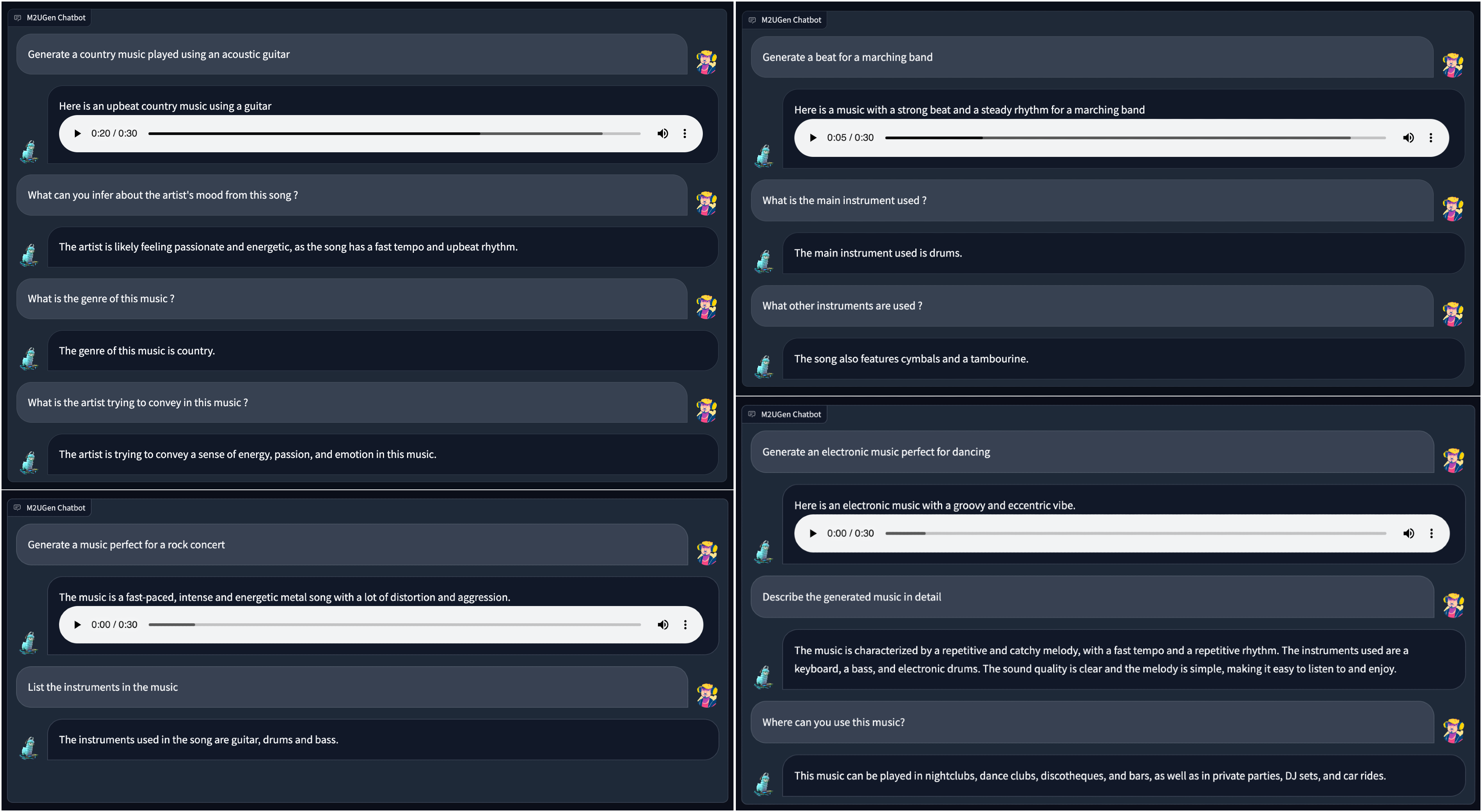}
    \caption{\textbf{Text-To-Music Generation and Understanding:} The MuMu-LLaMA model is capable of generating music from text prompts and also answering questions regarding the generated music.}
    \label{fig:T2M+MU}
\end{figure*}

\begin{figure*}[!t]
    \centering
    \includegraphics[width=\textwidth]{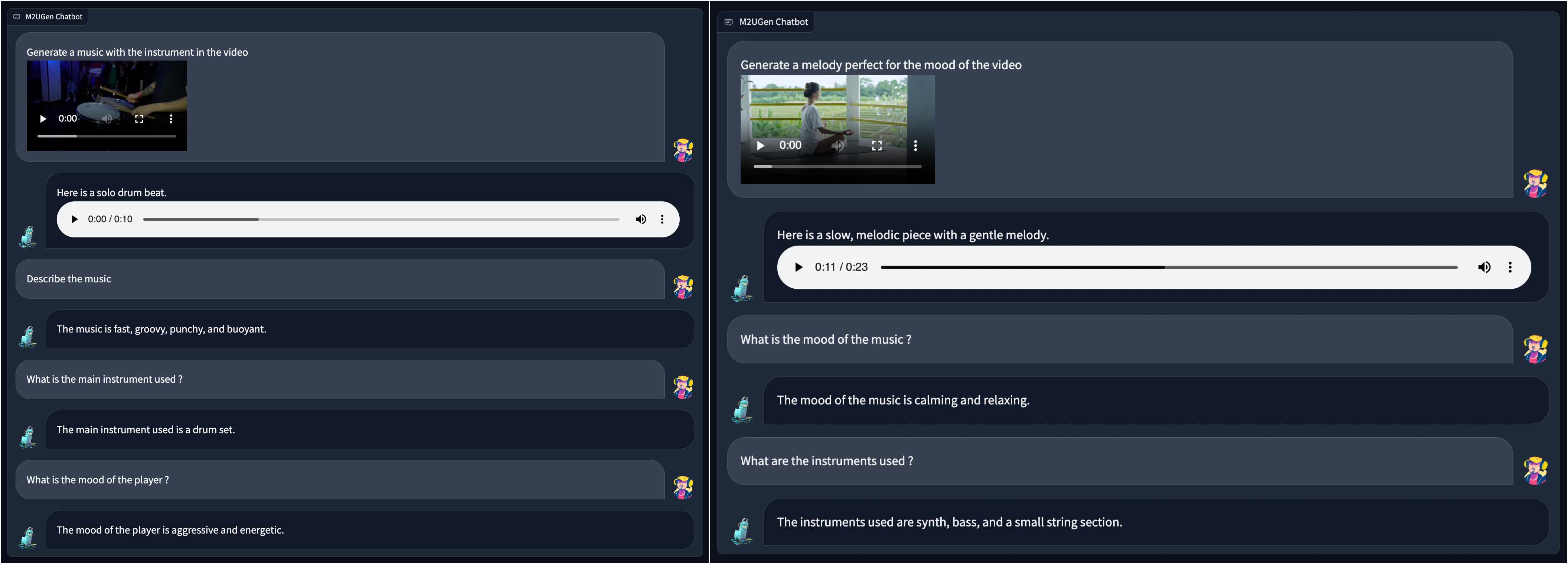}
    \caption{\textbf{Video-To-Music Generation and Understanding:} The MuMu-LLaMA model is capable of generating music for videos and also answering questions regarding the generated music.}
    \label{fig:V2M+MU}
\end{figure*}

\begin{figure*}[!t]
    \centering
    \includegraphics[width=0.9\textwidth]{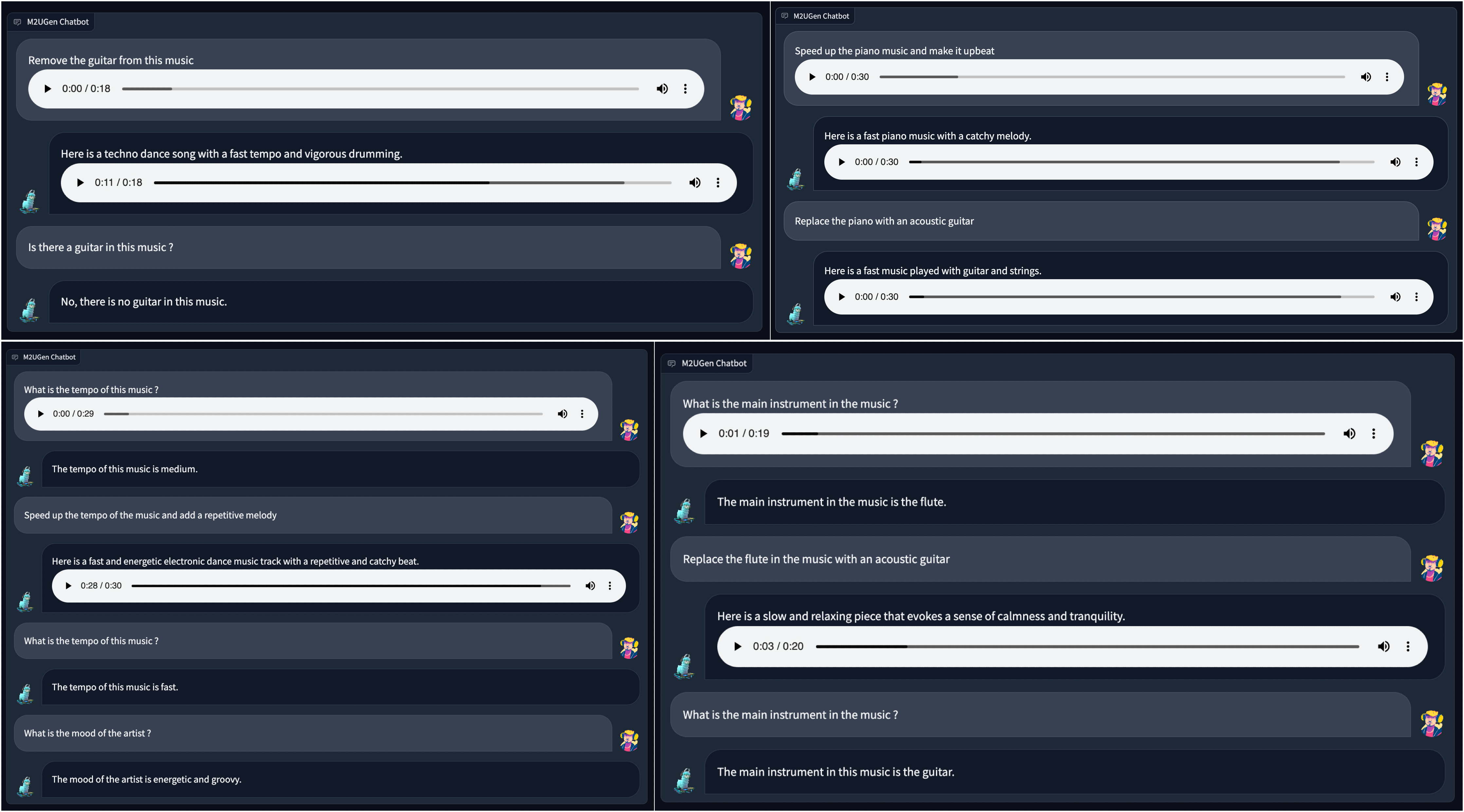}
    \caption{\textbf{Music Editing and Understanding:} The MuMu-LLaMA model is capable of editing input music based on natural language prompts and also answering questions regarding the generated music.}
    \label{fig:ME+MU}
\end{figure*}

\begin{figure*}[!t]
    \centering
    \includegraphics[width=0.9\textwidth]{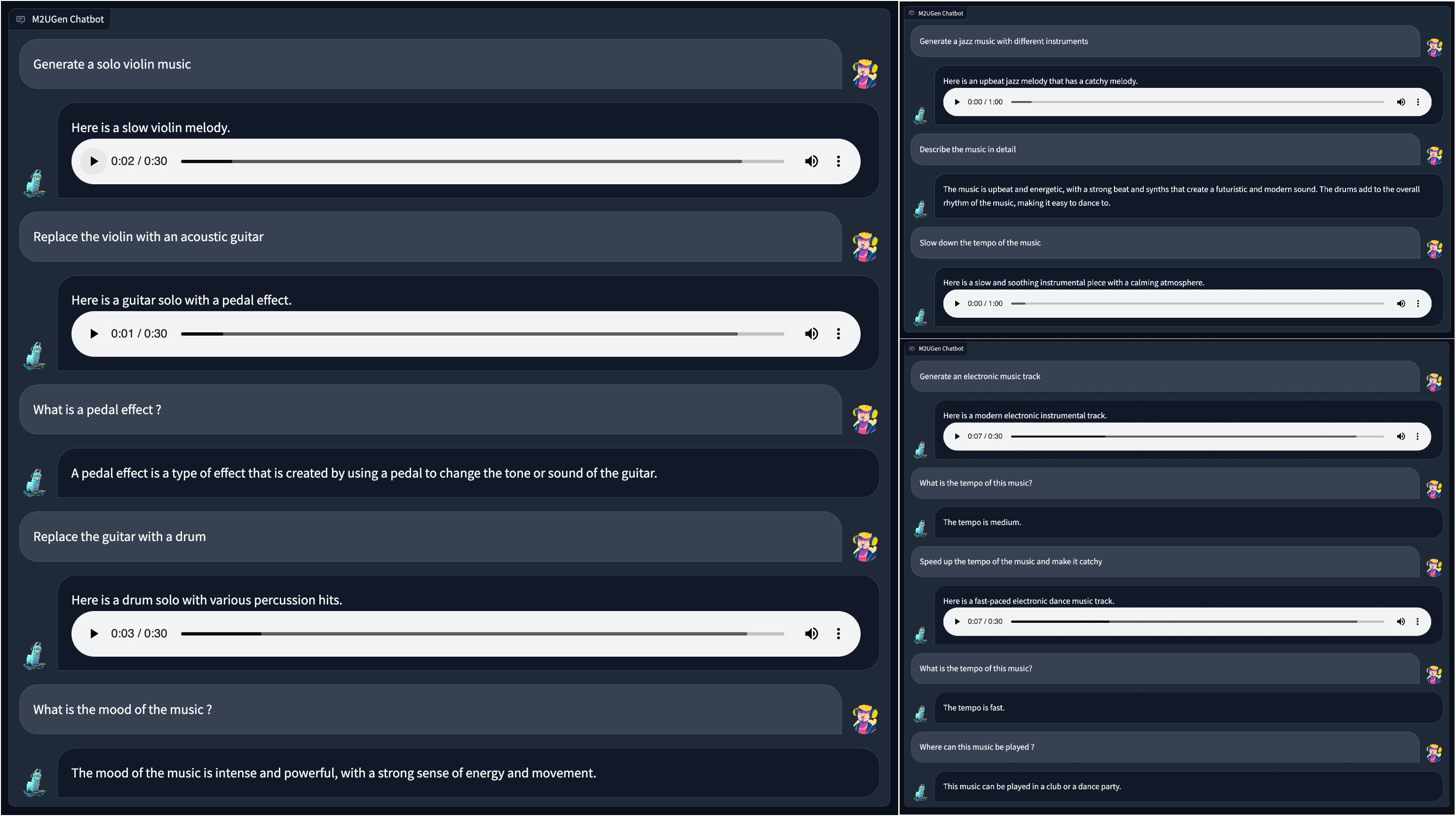}
    \caption{\textbf{Text-To-Music Generation and Understanding + Music Editing:} The MuMu-LLaMA model is capable of generating music from text prompts, answering questions regarding the generated music and also editing the generated music using Natural Language prompts.}
    \label{fig:T2M+MU+ME}
\end{figure*}

\begin{figure*}[!t]
    \centering
    \includegraphics[width=\textwidth]{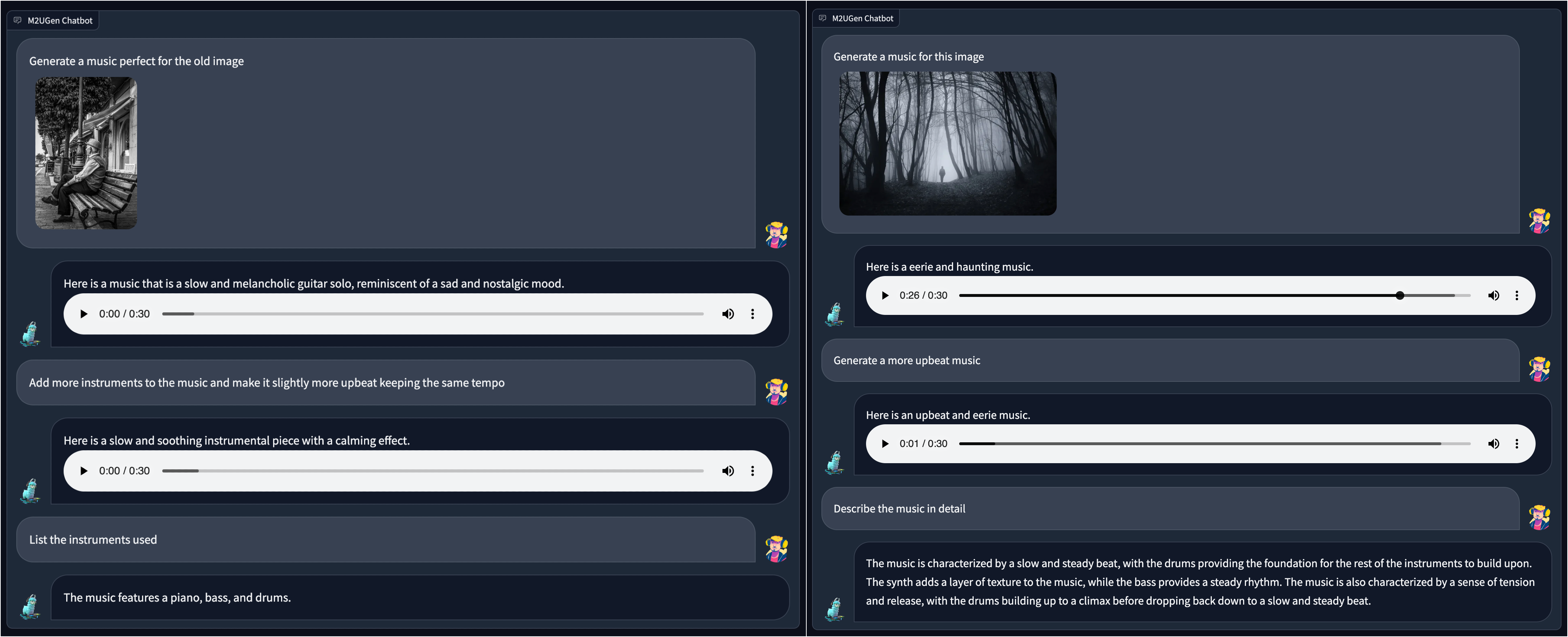}
    \caption{\textbf{Image-To-Music Generation and Understanding + Music Editing:} The MuMu-LLaMA model is capable of generating music for images, answering questions regarding the generated music and also editing the generated music using Natural Language prompts.}
    \label{fig:I2M+MU+ME}
\end{figure*}

\begin{figure*}[!t]
    \centering
    \includegraphics[width=0.9\textwidth]{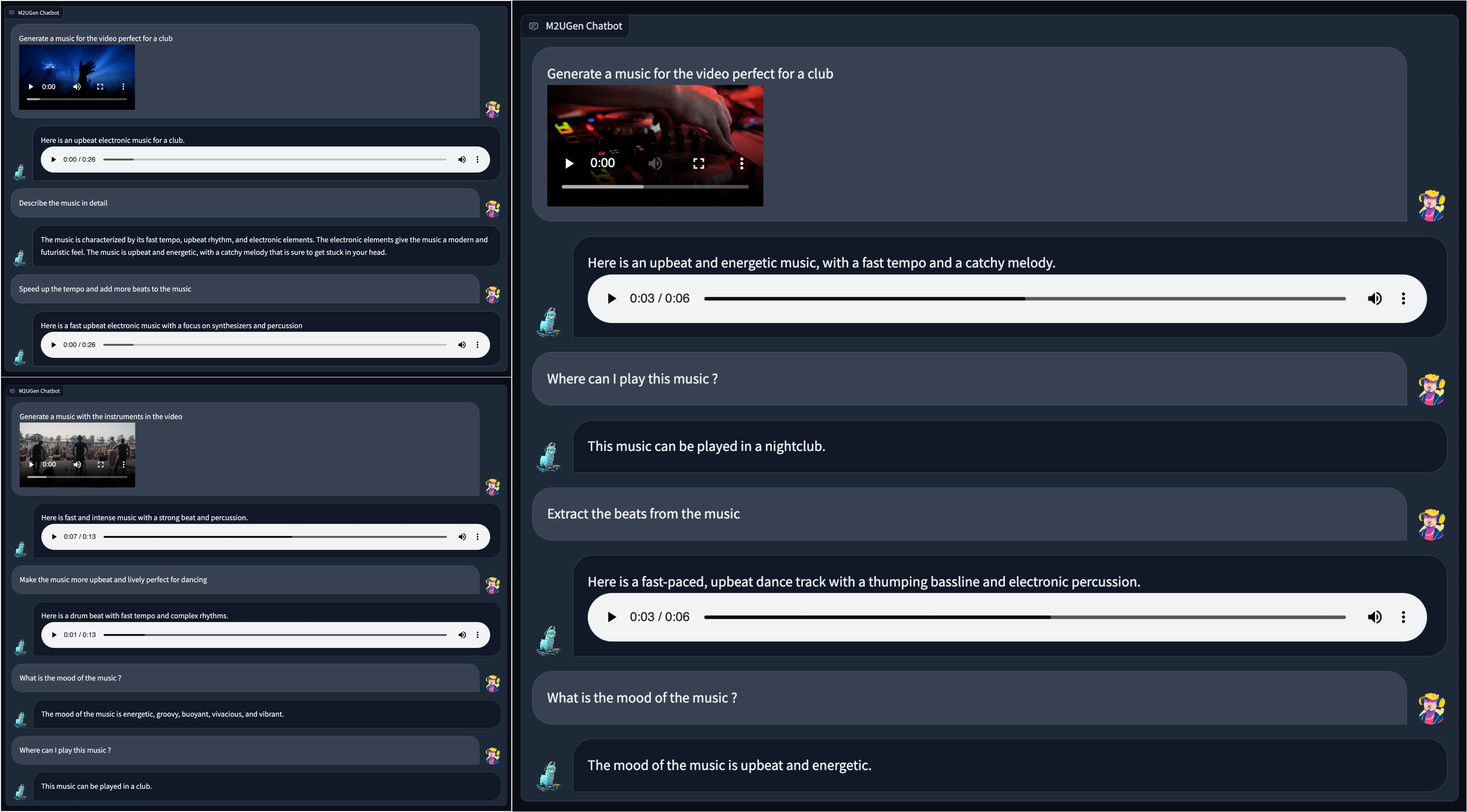}
    \caption{\textbf{Video-To-Music Generation and Understanding + Music Editing:} The MuMu-LLaMA model is capable of generating music for videos, answering questions regarding the generated music and also editing the generated music using Natural Language prompts.}
    \label{fig:V2M+MU+ME}
\end{figure*}

Figures \ref{fig:T2M+MU}, \ref{fig:I2M+MU}, and \ref{fig:V2M+MU} showcase the MuMu-LLaMA model's ability to generate music directly from textual prompts and draw inspiration from images and videos, both with and without textual guidance. Figure \ref{fig:ME+MU} exemplifies MuMu-LLaMA's proficiency in music editing guided by natural language prompts. Additionally, Figures \ref{fig:T2M+MU+ME}, \ref{fig:I2M+MU+ME}, and \ref{fig:V2M+MU+ME} illustrate the utilization of MuMu-LLaMA's editing capabilities to further refine music generated from different modalities. Collectively, the MuMu-LLaMA model proves to be a robust framework for Music Understanding, Generation, and Editing.
\end{document}